\documentclass[a4paper,10pt,oneside]{article}

\usepackage[english]{babel}
\usepackage[utf8]{inputenc}
\DeclareUnicodeCharacter{2010}{-}
\usepackage[T1]{fontenc}

\usepackage{lmodern}

\usepackage[hidelinks]{hyperref}

\usepackage{amsmath,amsfonts,amssymb,mathtools,mathrsfs,dsfont,mathabx}
\usepackage{accents}
\usepackage[mathscr]{euscript}
\usepackage{upgreek}

\usepackage[left=2cm,right=2cm,top=2cm,bottom=3cm]{geometry}

\usepackage{graphicx}
\graphicspath{{../../submission/}}

\usepackage{xcolor}

\usepackage{array,booktabs,multirow,makecell,tabu,colortbl}

\usepackage{enumerate}

\usepackage{subcaption}

\usepackage[numbers,sort&compress]{natbib}
\usepackage{doi}

\newcommand{\ab}{\boldsymbol{a}}
\newcommand{\bb}{\boldsymbol{b}}
\newcommand{\cb}{\boldsymbol{c}}

\newcommand{\eb}{\boldsymbol{e}}

\newcommand{\kb}{\boldsymbol{k}}

\newcommand{\pb}{\boldsymbol{p}}

\newcommand{\ub}{\boldsymbol{u}}
\newcommand{\vb}{\boldsymbol{v}}
\newcommand{\xb}{\boldsymbol{x}}

\newcommand{\zb}{\boldsymbol{z}}

\newcommand{\rhob}{\boldsymbol{\rho}}

\newcommand{\Ab}{\boldsymbol{A}}
\newcommand{\Bb}{\boldsymbol{B}}

\newcommand{\Eb}{\boldsymbol{E}}
\newcommand{\Fb}{\boldsymbol{F}}

\newcommand{\Jb}{\boldsymbol{J}}

\newcommand{\Pb}{\boldsymbol{P}}

\newcommand{\Xb}{\boldsymbol{X}}

\newcommand{\Zb}{\boldsymbol{Z}}

\newcommand{\Rbb}{\mathbb{R}}

\newcommand{\kpar}{k_{\|}}

\newcommand{\ppar}{p_{\|}}

\newcommand{\Fpar}{F_{\|}}

\newcommand{\Ppar}{P_{\|}}

\newcommand{\ctxt}{\text{c}}

\newcommand{\etxt}{\tn{e}}

\newcommand{\itxt}{\tn {i}}

\newcommand{\mtxt}{\text{m}}

\newcommand{\ptxt}{\text{p}}

\newcommand{\stxt}{\text{s}}

\newcommand{\Btxt}{\text{B}}

\newcommand{\Ttxt}{\text{T}}

\newcommand{\fhat}{\widehat{f}}

\newcommand{\nhat}{\widehat{n}}

\newcommand{\that}{\widehat{t}}

\newcommand{\vhat}{\widehat{v}}
\newcommand{\xhat}{\widehat{x}}

\newcommand{\Bhat}{\widehat{B}}

\newcommand{\Ehat}{\widehat{E}}

\newcommand{\That}{\widehat{T}}

\newcommand{\pbar}{\widebar{p}}

\newcommand{\Gbar}{\widebar{G}}
\newcommand{\Hbar}{\widebar{H}}

\newcommand{\Lbar}{\widebar{L}}

\newcommand{\Obar}{\widebar{O}}

\newcommand{\Xbbar}{\widebar{\boldsymbol{X}}}
\newcommand{\Zbbar}{\widebar{\boldsymbol{Z}}}

\newcommand{\Gcal}{\mathcal{G}}
\newcommand{\Hcal}{\mathcal{H}}

\newcommand{\Jcal}{\mathcal{J}}
\newcommand{\Lcal}{\mathcal{L}}

\newcommand{\Lbcal}{\boldsymbol{\mathcal{L}}}

\newcommand{\Qbcal}{\boldsymbol{\mathcal{Q}}}

\newcommand{\Zbcal}{\boldsymbol{\mathcal{Z}}}

\newcommand{\RR}{\mathbb{R}}

\newcommand{\gab}{\boldsymbol{\gamma}}

\newcommand{\rhb}{\boldsymbol{\rho}}

\newcommand{\thbar}{\widebar{\theta}}

\newcommand{\rhobbar}{\widebar\rhb}
\newcommand{\Bstar}{B^{*}}
\newcommand{\Bbstar}{\Bb^{*}}

\newcommand{\Bstarpari}{B^{*}_{\|\itxt}}
\newcommand{\Bstarpare}{B^{*}_{\|\etxt}}
\newcommand{\Nabla}{\boldsymbol{\nabla}}
\newcommand{\nablapar}{\nabla_\parallel}
\newcommand{\nablaperp}{\Nabla_\perp}

\newcommand{\wt}{\widetilde}
\newcommand{\dd}{\tn{d}}
\newcommand{\de}{\partial}

\newcommand{\upmubar}{\widebar{\upmu}}
\newcommand{\upmuhat}{\widehat{\upmu}}

\newcommand{\onehalf}{\frac{1}{2}}
\newcommand{\threehalves}{\frac{3}{2}}
\newcommand{\fivehalves }{\frac{5}{2}}
\newcommand{\sevenhalves}{\frac{7}{2}}
\newcommand{\ldot}[1]{\accentset{\mbox{\Large$.$}}{#1}}

\newcommand{\be}{\begin{equation}}
\newcommand{\ee}{\end{equation}}
\newcommand{\gavg}[1]{\left\langle #1 \right\rangle}
\newcommand{\dt}[1]{\frac{\dd #1}{\dd t}}

\newcommand{\der }[2]{\frac{\dd #1}{\dd #2}}
\newcommand{\pder}[2]{\frac{\partial #1}{\partial #2}}
\let\eps\varepsilon
\let\tn\textnormal
\let\wh\widehat
\let\para\parallel
\let\pa\partial

\title{Gyrokinetic theory with polynomial transforms: \\ a model for ions and electrons
in maximal ordering}

\author{
Edoardo Zoni$^{\,a,b}$, Stefan Possanner$^{\,a,b}$ \\[5mm]
{\small $^a$ {\it  Max-Planck-Institut für Plasmaphysik, Boltzmannstraße 2, 85748 Garching }} \\
{\small $^b$ {\it  Technische Universität München, Zentrum Mathematik, Boltzmannstraße 3, 85748 Garching }}}

\date{}

\begin{document}

\maketitle

\begin{abstract}
We propose a novel derivation of the gyrokinetic field-particle Lagrangian for non-collisional
ion-electron plasmas in a magnetic background with strong variations (maximal ordering).
Our approach follows the two-step reduction process, where the guiding-center coordinate
transformation is followed by the gyrocenter coordinate transformation in the single-particle
phase space. For the first time both steps are addressed within a unique methodology,
based on near-identity coordinate transformations constructed as polynomial transforms.
These are well-defined transformations composed of a finite number of terms
that are linear and algebraic with respect to the generating functions. The derivation
is carried out in a fully non-dimensional framework, based on parameters
governing the magnetic fusion experiments ASDEX Upgrade and ITER. Our method leads
to a gyrokinetic Vlasov-Maxwell model for ions and electrons, derived without the
use of Lie perturbation methods. It is found that, based on the employed ordering,
curvature terms such as the gyro-gauge term and the Ba\~nos drift appear at first
order in the ion Hamiltonian, whereas ion polarization terms appear only at second
order. By contrast, curvature terms are absent from the first-order electron Hamiltonian,
where instead magnetic flutter plays a role.
\end{abstract}

\section{Introduction}
Gyrokinetics is one of the major frameworks used in theoretical and numerical studies
of low-frequency turbulence in magnetized fusion plasmas \citep{Garbet2010,Krommes2012}.
Gyrokinetic models are based on a change of coordinates in particle phase space that
separates fast from slow dynamics, namely cyclotron motion from drift motion. The
idea is to derive a reduced set of dynamical equations that contain enough information
for an adequate description of low-frequency phenomena in the plasma. As a result, the
dimensionality of the problem is reduced and so is the computational cost in numerical
simulations.

In mathematical terms, gyrokinetic theory can be considered as the asymptotic
study of the Vlasov-Maxwell
model with strong magnetic background in the quasi-neutral regime. This requires a suitable reformulation
of the equations such that the corresponding asymptotic limit can
be carried out in a meaningful way. The procedure can be understood in the context
of averaging the characteristics (or Lagrangian paths) of the Vlasov equation, as
described first in \citep{Northrop1963}. Rigorous mathematical accounts can be found,
for example, in \citep{FrenodSonnen2001,Bostan2010,Chartier2016}. A first attempt
to average the Vlasov-Maxwell system taking into account the self-consistent interaction
between plasma particles and electromagnetic fields can be found in \citep{FriemanChen1982}.

From another point of view, the studies of \citep{Littlejohn1979,Dubin1983,Hahm1988,Brizard1989}
laid the foundation of ``structure-preserving'' gyrokinetic theory. These works
first showed that averaging could be carried out on the level of the variational
principle, by transforming the particle Lagrangian (or its corresponding Poisson
bracket structure). This strategy has the advantage of preserving symmetries of the
plasma equations during the process of averaging, in particular their Hamiltonian
structure: the averaged equations
exactly conserve averaged versions of the true constants of the motion, such
as energy or momentum. Most of the gyrokinetic simulation codes \citep{ORB5,
GENE,GYSELA,XGC} are based on structure-preserving gyrokinetic models and often show
improved stability and accuracy. The derivation of gyrokinetic models has been discussed
and reviewed extensively in \citep{Brizard2007,Krommes2012}, and references therein.
The prevalent methodology is based on Lie transform perturbation theory, as presented
for example in \citep{Cary1981,Cary1983}, and most of the recent derivations have
been carried out in this framework, as in \citep{Tronko2016,Brizard2017,Tronko2017a,
Tronko2017b,Tronko2018}.

This work is motivated by the need for an easier access to gyrokinetic theory, without
having to rely extensively on Lie perturbation methods. Such methods, despite being
mathematically elegant, are formulated in the language of differential geometry and
may thus prevent readers from focusing on the essence of the gyrokinetic reduction.
Here, we propose a different method which is inspired by the guiding-center theory
of Littlejohn \citep{Littlejohn1983}. A similar approach has been suggested recently
in \citep{Scott2017} and worked out in the long-wavelength regime of gyrokinetics in
\citep{Possanner2018}. In this work we extend the methodology presented in \citep{Possanner2018}
to address the description of turbulence on the microscopic scale of plasmas, such as the ion Larmor radius. As in Lie transform perturbation theory,
our method is also based on near-identity phase-space coordinate transformations.
However, we propose to construct such transformations in a different way. More precisely,
our phase-space coordinate transformations are defined as polynomials of finite degree
in powers of a given perturbation parameter, hence the name ``polynomial transforms''.
The coefficients of such polynomials are the so-called generating functions (or generators)
and represent the degrees of freedom that allow us to separate fast and slow scales.
This ansatz is conceptually simpler than working with Lie transforms, which are asymptotic
series (thus not necessarily convergent) constructed as products of operator exponentials
which feature Lie derivatives along the generating vector fields. In this work we show that 
structure-preserving gyrokinetic equations can be derived without the use of such
a complex mathematical machinery. 

We follow the strategy of the two-step gyrokinetic reduction, where the guiding-center
coordinate transformation is followed by the gyrocenter coordinate transformation
\citep{Tronko2018}. To derive the reduced equations we apply polynomial transforms
in each of the two steps, leading to a unified methodology for the complete reduction process.
Moreover, in the spirit of asymptotic analysis our derivation
is carried out in a fully non-dimensional framework. The perturbation parameter $\eps\ll 1$
in our near-identity coordinate transformations is identified by
a rigorous normalization of the Vlasov-Maxwell model. Our ordering in powers of $\eps$
is then based on assumptions derived from realistic physical scenarios relevant for
existing and future fusion experiments, such as the Tokamaks ASDEX Upgrade and ITER.
In this way we clearly separate the physical assumptions (ordering) from the mathematical
model reduction (averaging with polynomial transforms). Our methodology based on
a priori normalization of the physical equations allows us to formulate a gyrokinetic
theory for both ions and electrons within the same physical scenario and assumptions.

This paper is organized as follows. Section \ref{sec_model} introduces the basic
equations of the Vlasov-Maxwell model for a non-collisional magnetized plasma, including
a field-theoretic Lagrangian formulation. Section \ref{sec_normalization} defines
the normalization scheme used for the purpose of non-dimensionalization and derives
an ordering pattern based on physical considerations. Section \ref{sec_results} describes
in detail the gyrokinetic reduction procedure, leading from physical phase-space coordinates
to gyrocenter coordinates, and outlines the main results of this work, namely gyrokinetic
Lagrangians for both ions and electrons in maximal ordering, together with the corresponding
particle equations of motion and Maxwell's equations. Appendix \ref{app_variat}
reviews briefly the field-theoretic Lagrangian formulation of the Vlasov-Maxwell model.
Appendix \ref{app:gc} repeats
some calculations pertaining to the guiding-center transformation, presented in detail
in \citep{Possanner2018}. Appendix \ref{app_proofs} contains the calculations pertaining
to the derivation and proof of our main results for the gyrocenter transformation,
namely Propositions 1-3 for ions and Propositions
4-6 for electrons.

\section{The Vlasov-Maxwell model}
\label{sec_model}
We consider a non-collisional plasma composed of ions and electrons described in
terms of the distribution functions $f_\stxt(t,\xb,\vb)$, where $\stxt$ denotes the
particle species, $t\in\Rbb^+$ denotes the time coordinate, and $(\xb,\vb)\in\Rbb^3
\times\Rbb^3$ are position and velocity coordinates in phase space. All equations
are written in SI units in the following. The distribution functions $f_\stxt$ obey
the non-collisional Vlasov equation
\be
\label{Vlasov}
\pder{f_\stxt}{t}+\vb\cdot\Nabla f_\stxt+\frac{q_\stxt}{m_\stxt}\Big(\Eb+\vb\times 
\Bb\Big)\cdot\pder{f_\stxt}{\vb}=0 \,,
\ee
where $q_\stxt$ and $m_\stxt$ denote the particle charge and mass, respectively.
The electromagnetic fields $\Eb(t,\xb)$ and $\Bb(t,\xb)$ satisfy Maxwell's equations
\begin{subequations}
\begin{align}
& \Nabla\cdot\Eb =\frac{\rho}{\eps_0} \quad && \tn{(Coulomb's law)} \,, \\[1mm]
& \Nabla\cdot\Bb =0\quad && \tn{(absence of free magnetic poles)} \,, \\[1mm]
& \Nabla\times\Eb=-\pder{\Bb}{t} \quad && \tn{(Faraday's law)} \,, \\
& \Nabla\times\Bb =\mu_0\Jb+\eps_0\mu_0\pder{\Eb}{t} \quad && \tn{(Amp\`ere-Maxwell's law)} \,,
\end{align}
\end{subequations}
where $\eps_0$ and $\mu_0$ denote the vacuum electric permittivity and the vacuum
magnetic permeability, respectively. The sources $\rho(t,\xb)$ and $\Jb(t,\xb)$
are expressed in terms of the distribution functions as
\begin{subequations}
\label{sources}
\begin{align}
\rho & =\sum_\stxt q_\stxt\,n_\stxt=\sum_\stxt q_\stxt\,\int\dd^3\vb\,f_\stxt(t,\xb,\vb) \,, \\
\Jb  & =\sum_\stxt q_\stxt\,n_\stxt \ub_\stxt=\sum_\stxt q_\stxt\,\int\dd^3\vb\,\vb\,f_\stxt(t,\xb,\vb) \,.
\end{align}
\end{subequations}
The derivation of the Vlasov-Maxwell system \eqref{Vlasov}-\eqref{sources} from an
action principle was recognized first by \citep{Low1958}. Denoting by $\phi(t,\xb)$
and $\Ab(t,\xb)$ the electric scalar potential and the magnetic vector potential
associated to the electric and magnetic fields via $\Eb=-\Nabla\phi
-\de\Ab/\de t$ and $\Bb=\Nabla\times\Ab$, Low's action principle reads
\be
\label{Low:action}
\delta\int_{t_0}^{t_1}\dd t\left(L_\ptxt+L_\text{EM}\right)=0 \,,
\ee
where $\delta$ denotes the functional derivative and the Lagrangian is the sum of the electromagnetic free-field Lagrangian
\be
\label{def:Lf}
L_\text{EM}=\frac{\eps_0}{2}
\int\dd^3\xb\,\left|\Nabla\phi+\pder{\Ab}{t}\right|^2-\frac{1}{2\mu_0}\int\dd^3\xb\,
|\Nabla\times\Ab|^2 \,,
\ee
and the particle Lagrangian
\be
\label{def:Lp}
L_\ptxt=\sum_\stxt\int\dd^3\xb_0\,\dd^3\vb_0\,
f_{\stxt}(t_0,\xb_0,\vb_0)\,L_\stxt \,.
\ee
Here, $L_\stxt$ denotes the single-particle Lagrangian for the respective particle
species, defined
on the tangent bundle of the single-particle phase space. In the phase-space coordinates
$(\xb,\vb)$, the single-particle Lagrangian $L_\stxt$ reads
\be
\label{def:Ls}
L_\stxt=\left(m_\stxt\vb+q_\stxt\Ab\right)\cdot\ldot\xb-H_\stxt 
\quad \text{with} \quad
H_s:=m_\stxt|\vb|^2/2+q_\stxt\phi \,.
\ee
Hence, $L_\stxt$ depends implicitly on the potentials $\phi$ and $\Ab$ and describes
the self-consistent interaction between plasma particles and electromagnetic fields.

The variational principle \eqref{Low:action} leads to the characteristics of the
Vlasov equation by computing variations of $L_\stxt$ with respect to single-particle
trajectories $(\xb(t),\vb(t))$, to Coulomb's
law by computing variations with respect to $\phi$, and to Amp\`ere-Maxwell's law
by computing variations with respect to $\Ab$. We refer to Appendix \ref{app_variat}
for more details. 
Only the non-homogeneous Maxwell's equations, featuring source terms coupling to
the plasma particles, can be derived from the variational principle. With appropriate
initial/boundary conditions, this results in a well-posed system for $(f_\stxt,\phi,\Ab)$,
which describes the self-consistent interaction between plasma particles and electromagnetic fields.

An important aspect in gyrokinetic theory is the separation of the electromagnetic
fields into background and fluctuating parts. In this work we assume that the magnetic
field consists of dynamic fluctuations added to a static background, while the electric
field consists only of dynamic fluctuations (without a static background):
\be
\label{def:B} 
\Bb(t,\xb)=\Bb_0(\xb)+\Bb_1(t,\xb) \,, \qquad
\Eb(t,\xb)=\Eb_1(t,\xb) \,.
\ee
A dynamic background electric field $\Eb_0(t,\xb)$ can be studied within the drift-kinetic
framework, as in \citep{Possanner2018}, but is neglected here in order to focus on 
aspects specific to the gyrokinetic regime. Similarly, the electromagnetic potentials
are written as
\be
\Ab(t,\xb)=\Ab_0(\xb)+\Ab_1(t,\xb) \,, \qquad
\phi(t,\xb)=\phi_1(t,\xb) \,,
\ee
so that $\Bb_0=\Nabla\times\Ab_0$. Therefore, in the variational principle \eqref{Low:action}
variations have to be computed with respect to the dynamic potentials
$\phi_1$ and $\Ab_1$.

\section{Normalization and ordering}
\label{sec_normalization}
The formulation of the Vlasov-Maxwell system as a perturbation problem requires the
non-dimensionalization (or scaling) of the physical equations and the subsequent
application of an ordering scheme that allows for a comparison of terms in relation
to an \emph{asymptotic parameter} $\eps \ll 1$. Since gyrokinetics is ultimately
the theory of low-frequency dynamics in strongly-magnetized plasmas, the perturbation
parameter is typically defined as the ratio between a characteristic ion turbulence
frequency $\wh\omega_\itxt$ and the ion cyclotron frequency $\omega_\text{ci}$:
\be
\label{def:eps2}
\eps=\frac{\wh\omega_\itxt}{\omega_\text{ci}} \,.
\ee
Gyrokinetic theory can then be considered formally as the asymptotic analysis of the Vlasov-Maxwell
model in the limit $\eps \to 0$. The procedure of non-dimensionalization and the
introduction of the scaling parameter $\eps$ in the equations are fundamental. Therefore,
we give here a detailed description of both steps, which typically are not treated
extensively in the gyrokinetic literature. We first introduce the generic normalization
of the Low action principle \eqref{Low:action} and then suggest an ordering scheme
that corresponds to a realistic physical scenario relevant for existing and future
fusion experiments, such as the Tokamaks ASDEX Upgrade and ITER. 

\subsection{Normalization of the Vlasov-Maxwell model}
\label{subsec:normVM}
In order to write the Vlasov-Maxwell model in non-dimensional form, we introduce 
reference scales (denoted by a hat) for time, length, and ion and electron velocities:
\be
\label{scales1}
t=\wh t\, t' \,, \qquad \xb=\xhat\,\xb' \,, \qquad
\vb=
\begin{dcases}
\vhat_\itxt\,\vb' & \text{ions} \,, \\
\vhat_\etxt\,\vb'' & \text{electrons} \,.
\end{dcases}
\ee
Here, the primed quantities $t', x', \vb'$ and $\vb''$ are non-dimensional and the
characteristic velocities for ions and electrons, $\vhat_\itxt$ and $\vhat_\etxt$,
can be chosen differently, which means that $\vb'$ and $\vb''$ are different velocity
coordinates. We then write the background magnetic field and its corresponding magnetic
vector potential as
\be \label{B0'}
\Bb_0(\xb)=\Bhat_0\,\Bb_0'\left(\frac{\wh x}{\ell_0}\xb'\right) \,,\qquad
\Ab_0(\xb)=\wh A_0\,\Ab_0'\left(\frac{\wh x}{\ell_0}\xb'\right) \,,
\ee
where $\ell_0=||\Nabla\Bb_0/B_0||^{-1}$ denotes the \emph{length scale} of the background
magnetic field. Here, the primed functions $\Bb_0'$ and $\Ab_0'$ are dimensionless,
of order $O(1)$ and with variations of order $O(1)$ in the limit $\eps\to 0$. In
particular, they only take non-dimensional (scaled) arguments. Choosing $\wh A_0
:=\wh B_0\ell_0$, we obtain ${\Bb_0'=\Nabla'\times\Ab_0'}$ in the scaled variables.
We remark that for a uniform background magnetic field the considerations for the
vector potential $\Ab_0$ are still valid ($\Ab_0=(\Bb_0\times\xb)/2$ in this case).
With regard to the dynamic fields $\Bb_1$ and $\Eb_1$, we denote their length and
time scales by $\ell_1$ and $\tau_1$, respectively. For the magnetic and electric
fluctuations (and their corresponding potentials) we write
\begin{subequations}
\begin{alignat}{2}
\label{B1'}
& \Bb_1(t,\xb)=\wh B_1\,\Bb_{1}'\left(\frac{\wh t}{\tau_1}t',\frac{\wh x}{\ell_1}\xb'\right) \,, \qquad
\Ab_1(t,\xb) && =\wh A_{1}\,\Ab_{1}'\left(\frac{\wh t}{\tau_1}t',\frac{\wh x}{\ell_1}\xb'\right) \,, \\[1mm]
& \Eb_1(t,\xb)=\wh E_1\,\Eb_{1}'\left(\frac{\wh t}{\tau_1} t',\frac{\wh x}{\ell_1}\xb'\right) \,, \qquad
\phi_1(t,\xb) && =\wh\phi_1\,\phi_{1}'\left(\frac{\wh t}{\tau_1} t',\frac{\wh x}{\ell_1}\xb'\right) \,.
\end{alignat}
\end{subequations}
Choosing $\wh A_1:=\wh B_1\ell_1$ and $\wh\phi_1:=\wh E_1\ell_1$, we obtain
\be
\label{E1'}
\Bb_{1}'=\Nabla'\times\Ab_{1}' \,, \qquad
\Eb_{1}'=-\Nabla'\phi_{1}'-\frac{\wh B_1\ell_1}{\wh E_1\tau_1}\pder{\Ab_{1}'}{t'} \,,
\ee
in the scaled variables. We remark that the amplitudes $\wh A_1$ and $\wh\phi_1$
depend on the length scale $\ell_1$ of the fluctuations. Therefore, if the sizes
of the field fluctuations $\wh B_1$ and $\wh E_1$ are fixed, small-scale fluctuations
are associated to small potentials, while large-scale fluctuations are associated
to large potentials. The size of the potentials, in turn, plays a role in the ordering
of terms in the particle Lagrangian, and thus in the overall asymptotic expansion.

In the Low action \eqref{Low:action} we normalize the electromagnetic free-field Lagrangian
\eqref{def:Lf} as $L_\tn{EM}=\wh n_\itxt\,k_\Btxt\,\wh T_\itxt\,\wh x^3\,L_\tn{EM}'$,
where $\wh n_\itxt$ denotes a reference ion density, $\wh T_\itxt$ a reference ion
temperature, and $k_\Btxt$ is the Boltzmann constant. Therefore, we obtain the non-dimensional
electromagnetic free-field Lagrangian
\be
\label{Lf'}
L_\tn{EM}'= 
\frac{\eps_0\,\wh E_1^2}{\wh n_\itxt\,k_\Btxt\,\wh T_\itxt}\frac{1}{2}
\int\dd^3\xb'\,\bigg|\Nabla'\phi_{1}'+\frac{\wh B_1\ell_1}{\wh E_1\tau_1}\,\pder{\Ab_{1}'}{t'}\bigg|^2 
-\frac{1}{\beta_\itxt}
\int\dd^3\xb'\,\bigg|\Nabla'\times\Ab_0'+\frac{\wh B_1}{\wh B_0}\,\Nabla'\times\Ab_{1}'\bigg|^2 \,,
\ee
where $\beta_\itxt=2\,\mu_0\,\wh n_\itxt\,k_\Btxt\,\wh T_\itxt/\wh B_0^2$ is the ion
plasma beta.
In order to normalize the single-particle Lagrangian \eqref{def:Ls}, we note that
$\ldot\xb$ represents three components of an element $(\ldot\xb,\ldot\vb)\in\Rbb^3\times\Rbb^3$
of the tangent space at $(\xb,\vb)\in\Rbb^3\times\Rbb^3$. Tangents $\ldot\xb$ have
units of velocity and are therefore normalized as $\ldot\xb=(\wh x/\,\wh t\,)\,\ldot\xb'$.
We then normalize the single-particle Lagrangian \eqref{def:Ls} as $L_\stxt=m_\stxt
\,\wh v_\stxt^2\,L_\stxt'$, obtaining the non-dimensional single-particle Lagrangian
\be
\label{Ls'}
L_\stxt'=\frac{\wh x}{\wh v_\stxt\,\wh t}\left[\vb'+\frac{q_\stxt\wh B_0\,\ell_0}{m_\stxt\wh v_\stxt}
\left(\Ab_0'+\frac{\wh B_1\ell_1}{\wh B_0\ell_0}\,\Ab'_{1}\right)\right]\cdot\ldot{\xb}'
-\frac{|\vb'|^2}{2}-\frac{q_\stxt\wh E_1\ell_1}{m_\stxt\wh v_\stxt^2}\phi_{1}' \,.
\ee
Finally, we normalize the particle Lagrangian of the Low action in the same way
as the electromagnetic free-field Lagrangian, namely as $L_\ptxt=\wh n_\itxt\,k_\Btxt\,
\wh T_\itxt\,\wh x^3\,L_\ptxt'$, obtaining the non-dimensional particle Lagrangian
\be
\label{Lp'}
L_\ptxt'=\sum_\stxt\frac{m_\stxt\,\wh v_\stxt^5\,\wh f_s}{\wh n_\itxt\,k_\Btxt\wh T_\itxt}
\int\dd^3 \xb_0'\,\dd^3\vb_0'\,f_{\stxt}'(t_0',\xb_0',\vb_0')\,L_\stxt' \,.
\ee
All dependent variables in the Low action are now expressed in terms of non-dimensional
functions. Therefore, the size of each term is determined only by the size of the
non-dimensional coefficients in front of it. Such coefficients are, in turn, determined
by the physical scenario under consideration, as we discuss in the next section.

\subsection{Physical scenario and ordering}
\label{sec:phys}
The normalization of the Vlasov-Maxwell system described in the previous section
is generic. For the purpose of deriving a set of gyrokinetic equations by asymptotic
methods we must quantify the size of the non-dimensional coefficients appearing in
the physical quantities of interest in powers of the perturbation parameter $\eps$.
This process is usually referred to as \emph{ordering}. Different orderings lead
to different perturbation theories and to reduced models with different physical
content. An ordering is thus the mathematical expression of a specific physical
scenario. Two such scenarios for magnetic confinement fusion experiments are listed
in Table \ref{tab:scenarios}.
\begin{table}
\begin{center}
\tabulinesep=2mm
\begin{tabu}{|c|c|c|c|c|c|c|}
\hline
\multicolumn{3}{|c|}{} & \multicolumn{2}{c|}{ASDEX Upgrade} & \multicolumn{2}{c|}{ITER} \\
\tabucline{4-}
\multicolumn{3}{|c|}{} & ions & electrons & ions & electrons \\
\hline
major radius & $R_0$ & $[\mtxt]$ & \multicolumn{2}{c|}{$1.6$} & \multicolumn{2}{c|}{$6.2$} \\
\hline
minor radius & $a$ & $[\mtxt]$ & \multicolumn{2}{c|}{$0.8$} & \multicolumn{2}{c|}{$2.0$} \\
\hline
toroidal magnetic field & $B_\Ttxt$ & $[\Ttxt]$ & \multicolumn{2}{c|}{$3.9$} & \multicolumn{2}{c|}{$5.3$} \\
\hline
average particle density & $\langle n_\stxt\rangle$ & $[\dfrac{10^{20}}{\mtxt^3}]$
& $2.0$ & $2.0$ & $1.0$ & $1.0$ \\
\hline
average thermal energy & $\langle T_\stxt\rangle$ & $[\dfrac{\textrm{keV}}{k_\Btxt}]$ 
& $8.7$ & $8.7$ & $8.0$ & $8.8$ \\
\hline
cyclotron frequency & $\omega_{\ctxt\stxt}$ & $[\textrm{Hz}]$ & $1.9\times10^8$
& $6.9\times10^{11}$ & $2.5\times10^8$ & $9.3\times10^{11}$ \\
\hline
thermal velocity & $\vhat_\stxt$ & $[\dfrac{\mtxt}{\stxt}]$ & $6.4\times10^5$
& $3.9\times10^7$ & $6.2\times10^5$ & $3.9\times10^7$ \\
\hline
thermal frequency & $\wh\omega_\stxt$ & $[\textrm{Hz}]$ & $8.0\times10^5$ & $4.9\times10^7$
& $3.1\times10^5$ & $2.0\times10^7$ \\
\hline
Larmor radius & $\rho_\stxt$ & $[\mtxt]$ & $3.4\times10^{-3}$ & $5.7\times10^{-5}$
& $2.4\times10^{-3}$ & $4.2\times10^{-5}$ \\
\hline
Debye length & $\lambda_\stxt$ & $[\mtxt]$ & $4.9\times10^{-5}$ & $4.9\times10^{-5}$
& $6.7\times10^{-5}$ & $7.0\times10^{-5}$ \\
\hline
\multicolumn{3}{|c|}{$\wh\omega_\stxt/\omega_{\ctxt\stxt}$} & $3.4\times10^{-3}$
& $5.7\times10^{-5}$ & $1.2\times10^{-3}$ & $2.1\times10^{-5}$ \\
\hline
\multicolumn{3}{|c|}{$v_\stxt^2/c^2$} & $4.6\times 10^{-6}$ & $1.7\times 10^{-2}$
& $4.3\times 10^{-6}$ & $1.7\times 10^{-2}$ \\
\hline
\multicolumn{3}{|c|}{$\lambda_\stxt^2/a^2$} & $3.7\times10^{-9}$ & $3.7\times10^{-9}$
& $1.1\times10^{-9}$ & $1.2\times10^{-9}$ \\
\hline
\multicolumn{3}{|c|}{$\beta_\stxt$}
& $1.2\times 10^{-2}$ & $1.2\times 10^{-2}$ & $2.4\times 10^{-3}$ & $3.2\times 10^{-3}$ \\
\hline
\end{tabu}
\end{center}
\caption{Physical scenarios for magnetic confinement fusion experiments: parameters
for the Tokamaks ASDEX Upgrade \citep{AUGweb} and ITER \citep{ITER2005}. Note that we choose $\Bhat_0=B_\Ttxt$,
$\nhat_\stxt=\langle n_\stxt\rangle$ and $\That_\stxt=\langle T_\stxt\rangle$.}
\label{tab:scenarios}
\end{table}
We choose as the characteristic length and time scales of observation the minor
radius $a$ and the inverse ion thermal frequency:
\be
\xhat:=a \,, \qquad \that^{-1}:=\wh\omega_\itxt \,.
\ee
The ion thermal frequency $\wh\omega_\itxt$, that is, the inverse of the time required
for an ion to travel the distance $a$, is close to the frequency of micro-turbulence
observed in Tokamaks \citep{Liewer1985,Wootton1990,Krommes2012}. By substituting
in \eqref{def:eps2} the values shown in Table \ref{tab:scenarios}, we then obtain
\begin{equation}
\eps\approx 10^{-3} \,.
\end{equation}
Measurements in Tokamaks have shown that fluctuation levels in turbulent plasmas satisfy \citep{Liewer1985,
Wootton1990,Brizard2007}
\be
\frac{\Bhat_1}{\Bhat_0}\sim\frac{\Ehat_1}{\Bhat_0\vhat_\itxt}\approx 10^{-3}\approx\eps \,.
\ee
Moreover, if we consider electrons and deuterium ions we have $q_\etxt/q_\itxt=-1$ and
\be
\label{ordering:end}
\frac{m_\etxt}{m_\itxt} \approx 2.7 \times 10^{-3}\approx\eps  \,.
\ee
We also notice an ordering
pattern in powers of $\eps$ in the normalized plasma parameters of Table \ref{tab:scenarios}
(last four non-dimensional parameters).
We then apply this ordering to the normalized Low action. By choosing
\be
\ell_0:=a \,, \qquad \ell_1:=\rho_\itxt \,, \qquad \tau_1^{-1}:=\wh\omega_\itxt \,,
\ee
we satisfy the maximal ordering $\ell_1/\ell_0=O(\eps)$ in the limit $\eps\to 0$.
More precisely, what we mean here is that the ratio $\ell_1/\ell_0$ is approximately
$\eps$ and therefore it can be considered as a function of order $O(\eps)$ if we
were taking the formal limit $\eps\to0$ (which we actually never take, as $\eps$
is a fixed number and cannot vanish).
The normalized background magnetic field \eqref{B0'}
becomes a function of $\xb'$, namely $\Bb_0'(\xb')$. On the other hand, since ${\ell_1/\xhat=O(\eps)}$, the fluctuating
electric and magnetic fields and their corresponding potentials become strongly-varying
functions of $\xb'$, namely $\Eb_1'(t',\xb'/\eps)$ and $\Bb_1'(t',\xb'/\eps)$.
More precisely, we assume the conventional gyrokinetic ordering for the fluctuations,
namely $|\kb_\perp|\rho_\itxt
=O(1)$ and $k_\parallel\rho_\itxt=O(\eps)$ in the limit $\eps\to 0$, where ${\kb:=
(k_\para,\kb_\perp)}$ denotes characteristic wave vectors parallel and perpendicular
to $\Bb_0$, respectively. Moreover, for the coefficient
appearing in \eqref{E1'} we have
\be
\frac{\Bhat_1\ell_1}{\Ehat_1\tau_1}=\frac{\Bhat_1\rho_\itxt\,\wh\omega_\itxt}{\Ehat_1}
=\frac{\wh\omega_\itxt}{\omega_{\ctxt\itxt}}\frac{\Bhat_1\,\vhat_\itxt}{\Ehat_1}=\eps\,
\frac{\Bhat_1}{\Bhat_0}\frac{\Bhat_0\,\vhat_\itxt}{\Ehat_1}\approx \eps \,,
\ee
which yields
\be
\Eb_{1}'=-\Nabla'\phi_{1}'-\eps\,\pder{\Ab_{1}'}{t'} \,.
\ee
Moreover, the non-dimensional coefficients in the normalized electromagnetic free-field
Lagrangian \eqref{Lf'} have the following sizes:
\be
\frac{\eps_0\Ehat_1^2 }{\nhat_\itxt k_\Btxt\That_\itxt}=\frac{\vhat_\itxt^2}{c^2}
\frac{\Ehat_1^2}{\Bhat_0^2 \vhat_\itxt^2}\frac{\Bhat_0^2}{\mu_0 \nhat_\itxt k_\Btxt \That_\itxt}
\approx \eps^3 \,, \qquad
\frac{1}{\beta_\itxt}
\approx \frac{1}{\eps} \,.
\ee
In the normalized particle Lagrangian \eqref{Lp'} we set $k_\Btxt\That_\etxt=k_\Btxt\That_\itxt$
and $\vhat_\stxt^3\fhat_\stxt=\nhat_\itxt$ to fix the characteristic size
of the distribution function. Therefore, in our ordering the normalized ion and electron
single-particle Lagrangians \eqref{Ls'} read
\begin{subequations}
\begin{align}
L'_\itxt &=  \bigg(\vb'
 + \frac{\Ab_0'}{\eps} + \eps\, \Ab_{1}' \bigg)
\cdot\ldot{\xb}' - \frac{|\vb'|^2}{2} - \eps \, \phi_{1}'\,,  \label{Li'}
 \\[2mm]
 L'_\etxt &=  \bigg( \sqrt{\frac{m_\etxt}{m_\itxt}} \vb'
 - \frac{\Ab_0'}{\eps} - \eps\,\Ab_{1}' \bigg)
\cdot\ldot{\xb}' - \frac{|\vb'|^2}{2} + \eps\, \phi_{1}'\,.  \label{Le'}
\end{align}
\end{subequations}
The only difference between ions and electrons, besides the sign in front of the
electromagnetic potentials due to the negative electron charge, is the factor
$\sqrt{m_\etxt/m_\itxt}\approx\sqrt{\eps}$ multiplying $\vb' \cdot \ldot \xb'$,
which defines an intermediate scale that is not an integer power of $\eps$.
As the final result of the ordering procedure, we obtain the normalized Low action principle
\be \label{Low'}
\delta \int_{t_0'}^{t_1'} (L_\tn{p}' + L_\tn{EM}')\, \tn d t' = 0\,,
\ee
with the Lagrangians given by (omitting the primes for a simpler notation)
\begin{subequations}
\begin{align}
  L_\tn{p}' &=  \int\dd^6\zb_0\,
  f_\itxt(t_0,\zb_0) \, \left[  \bigg( \vb
 + \frac{\Ab_0}{\eps} + \eps\, \Ab_{1} \bigg)
\cdot\ldot\xb
- \frac{|\vb|^2}{2} - \eps \, \phi_{1} \right]
 \\[1mm]
 & + \int\dd^6\zb_0\,
 f_\etxt(t_0,\zb_0)\,\left[ \bigg( \sqrt{\eps}\, \vb
 - \frac{\Ab_0}{\eps} - \eps\,\Ab_{1} \bigg)
\cdot\ldot\xb
- \frac{|\vb|^2}{2} + \eps\, \phi_{1} \right]
 \,,  \nonumber
 \\[3mm]
 L_\tn{EM}' &= \frac{\eps^3}{2} \int\dd^3 \xb\, \left| \Nabla \phi_{1} + \eps\, \pder{\Ab_{1}}{t} 
\right|^2 
 - \frac{1}{2\,\eps} \int\dd^3\xb\, \left| 
\Nabla\times\Ab_0 + \eps\, \Nabla\times\Ab_{1} \right|^2 \,.
\end{align}
\end{subequations}
Taking variations with respect to particle trajectories, 
$\phi_{1}$ and $\Ab_{1}$ in \eqref{Low'} leads to the 
following normalized Vlasov-Maxwell equations (again omitting the primes):
\begin{subequations}
\label{LT}
\begin{align}
 &\pder{f_\itxt}{t} + \vb\cdot\Nabla f_\itxt + \left[\Eb_{1} + \vb\times \left( \frac{\Bb_0}{\eps} 
+ \Bb_{1} \right) \right]\cdot\pder{f_\itxt}{\vb} = 0\,,
 \\[1mm]
 \sqrt{\eps}\,&\pder{f_\etxt}{t} + \vb\cdot\Nabla f_\etxt - \left[\Eb_{1} + \vb\times 
\frac{1}{\sqrt{\eps}}\left( \frac{\Bb_0}{\eps} + \Bb_{1} \right) \right]\cdot\pder{f_\etxt}{\vb} = 
0\,,
 \\[1mm]
 &\eps^2\,\Nabla\cdot\Eb_{1} = \rho \,, \label{Coulomb_scaled} \\[2mm]
 &\Nabla\times \left(\frac{\Bb_0}{\eps} + \Bb_{1} \right) = \Jb + \eps^3\,\pder{\Eb_{1}}{t} \,,
\label{Ampere_scaled}
\end{align}
\end{subequations}
where the normalized charge and current densities are given by
\be
\label{rhoJ:norm}
\rho=\int\dd^3\vb\,f_\itxt-\int\dd^3\vb\,f_\etxt \,, \qquad
\Jb=\int\dd^3\vb\,\vb\,f_\itxt-\frac{1}{\sqrt{\eps}}\int\dd^3\vb\,\vb\,f_\etxt \,.
\ee
The factor $1/\sqrt{\eps}$ in front of the electron current density comes from the
different choice of scales for the ion and electron thermal velocities. From the two
Vlasov equations in \eqref{LT} we deduce the charge conservation law
\be \label{rhoconserve}
 \pder{\rho}{t} + \nabla \cdot \Jb = 0\,.
\ee
This is also a solvability condition for Maxwell's equations. Indeed, taking the
divergence of Amp\`ere-Maxwell's law in \eqref{Ampere_scaled}, recalling that
$\Eb_1(t,\xb/\eps)$ is strongly-varying in space, 
and inserting Coulomb's law from \eqref{Coulomb_scaled} yields \eqref{rhoconserve}.

The normalized variational principle \eqref{Low'}, or the normalized set of equations
\eqref{LT}, are a suitable starting point for our perturbation analysis of the Vlasov-Maxwell
system. Let us remark that the majority of gyrokinetic theories 
for micro-turbulence have been developed in a homogeneous background for the sake
of conceptual clarity. However, because curvature is important in magnetically confined
fusion plasmas (``neo-classical transport''), many of the state-of-the-art numerical
codes feature a model with slowly-varying magnetic background, corresponding to
$\Bb_0(\eps\,\xb)$ in normalized variables. Curvature terms then appear only at the
second order of the perturbation theory, and are often neglected for simplicity. In
this work, we develop a consistent theory in the maximal ordering, corresponding to
$\Bb_0(\xb)$ in normalized variables. This seems to be the scenario for current Tokamak
and Stellarator experiments, but is also interesting for Spheromaks and even for
the future large-scale Tokamak ITER. Reduced equations for smaller background
curvature can easily be deduced from our more general results in maximal ordering.

\section{Gyrokinetic reduction}
\label{sec_results}

The basic idea of gyrokinetic theory is to replace the exact trajectories of the
plasma particles by the trajectories of their \emph{gyrocenters}, which move on
the time scale of the thermal frequency $\wh\omega_\itxt$ or slower. The dynamics
occurring at scales faster than the cyclotron frequency $\omega_\text{ci}$ are ``averaged
out'' in the gyrocenter picture. However, some effects of the fast motion of gyration
are still present in form of drifts of the gyrocenters. In this section we make these
concepts more precise by analyzing the formal asymptotic limit $\eps\to0$ in the normalized
single-particle Lagrangians \eqref{Li'}-\eqref{Le'}. From the reduced Lagrangians
we then derive the gyrokinetic Vlasov equation for ions and electrons and define
gyrocenter charge and current densities with polarization corrections, thus coupling
plasma particles and electromagnetic fields from the normalized gyrocenter action
principle. Primes are omitted from now on, in order to increase readability.

Following \citep{Sugama2000}, we intend to replace the particle Lagrangian $L_\tn{p}$
in the Low action principle \eqref{Low'} by its gyrocenter representation $\Lcal_\tn{p}$:
\begin{equation*}
\int\dd^6\zb_0\,\left[f_\itxt(0,\zb_0)\,L_\itxt+f_\etxt(0,\zb_0)\,L_\etxt\right]\approx
\int\dd^6\Zbcal_0\left[B^*_{\para\itxt}F_\itxt(0,\Zbcal_0)\,\Lcal_\itxt+B^*_{\para\etxt}F_\etxt(0,\Zbcal_0)\,\Lcal_\etxt\right] \,,
\end{equation*}
where $\dd^6\Zbcal_0:=\dd^3\Xb_0\,\dd P_{\|0}\,\dd\mu_0\,\dd\Theta_0$ denotes the
measure in gyrocenter phase space, $F_\stxt$ denotes the gyrocenter distribution
function, $\Lcal_\stxt$ is the corresponding gyrocenter single-particle Lagrangian,
to be derived below, and $\Bstar_{\|\stxt}$ is the Jacobian determinant.
The gyrocenter coordinates are the gyrocenter position $\Xb$,
the gyrocenter parallel momentum $\Ppar$, the gyrocenter magnetic moment $\mu$
and the gyro-angle $\Theta$. The single-particle dynamics in the new coordinates
is such that the time evolution of the gyro-angle $\Theta$ is decoupled from the
rest of the coordinates, leading to a closed system of equations for the ``slow''
variables $(\Xb,\Ppar)$, where $\mu$ is a constant of the motion. The slow system
represents the averaged dynamics (see \citep{Possanner2018} for details).

The phase-space coordinate transformation $\Zbcal:=(\Xb,\Ppar,\mu,\Theta)\mapsto(\xb,\vb)$
with Jacobian determinant denoted by $B^*_\|$ is the central object of gyrokinetic theory.
It is usually derived in terms of (canonical) Lie transforms of the fundamental one-form
associated to the single-particle Lagrangian $L_\stxt$ \citep{Brizard1989,Hahm1988,Tronko2018}.
Despite being an elegant mathematical framework, Lie transform perturbation theory
introduces many formal complications, which seem not to be strictly necessary for
averaging. In this work we replace Lie transforms with polynomials of finite degree
in $\eps$, algebraic in the generating functions. We show that also with this different
ansatz for the phase-space near-identity coordinate transformation it is possible
to remove the gyro-angle dependence from the single-particle Lagrangian up to the
desired order in $\eps$, without changing its symplectic part (and thus the Jacobian
$B^*_\para$ of the coordinate transformation). Our polynomial transforms are well-defined
coordinate transformations (locally invertible), in contrast to the asymptotic series
in Lie transform perturbation theory, where it is difficult to prove convergence
and existence of the transforms. It is our hope that the simpler derivation based
on polynomial transforms will enable more rigorous mathematical studies of gyrokinetic
theory in the future.

\subsection{Preliminary transformations}
The phase-space coordinate transformation $\Zbcal\mapsto(\xb,\vb)$ is a composition
of several coordinate changes, which are summarized in Tables \ref{tab_super_ions}
and \ref{tab_super_electrons} for ions and electrons, respectively.
\begin{table}
\begin{center}
\tabulinesep=2mm
\begin{tabu}{|c|c|}
\hline
$\begin{gathered}
\text{physical coordinates} \\[2pt] (\xb,\vb)
\end{gathered}$ & \\
\hline
$\begin{gathered}
\text{Hamiltonian picture} \\[2pt] (\xb,\pb)
\end{gathered}$
& $\pb:=\vb+\eps\,\Ab_1$ \\
\hline
$\begin{gathered}
\text{angle coordinates} \\[2pt] (\xb,\ppar,\upmu,\theta)
\end{gathered}$ &
$\begin{alignedat}{2}
& \ppar  && :=\pb\cdot\bb_0 \\[1mm]
& \upmu  && :=\frac{1}{2B_0}\left|\bb_0\times\pb\times\bb_0\right|^2 \\
& \theta && :=\arctan\left(\frac{\pb\cdot\eb_1}{\pb\cdot\eb_2}\right)
\end{alignedat}$ \\[-1mm]
\hline
$\begin{gathered}
\text{guiding-center coordinates} \\[2pt] \Zbbar:=(\Xbbar,\pbar_\|,\upmubar,\thbar)
\end{gathered}$ &
$\begin{alignedat}{5}
& \xb && :=\Xbbar && +\eps\,\bar\rhob_{1\itxt} && +\eps^2\bar\rhob_{2\itxt} &&
+\eps^3\bar\rhob_{3\itxt}+\eps^4\bar\rhob_{4\itxt} \\[5pt]
& \ppar && :=\pbar_\| && +\eps\,\Gbar_{1\itxt}^\| && +\eps^2\Gbar_{2\itxt}^\| && +\eps^3\Gbar_{3\itxt}^\| \\[5pt]
& \upmu && :=\upmubar && +\eps\,\Gbar_{1\itxt}^\upmu && +\eps^2\Gbar_{2\itxt}^\upmu && +\eps^3\Gbar_{3\itxt}^\upmu \\[5pt]
& \theta && :=\thbar && +\eps\,\Gbar_{1\itxt}^\Theta && +\eps^2\Gbar_{2\itxt}^\Theta && +\eps^3\Gbar_{3\itxt}^\Theta
\end{alignedat}$ \\
\hline
$\begin{gathered}
\text{preliminary gyrocenter coordinates} \\[2pt] \Zb:=(\Xb,\Ppar,\upmuhat,\Theta)
\end{gathered}$ &
$\begin{alignedat}{5}
& \Xbbar && :=\Xb && && +\eps^2\rhob_{2\itxt} && +\eps^3\rhob_{3\itxt}+\eps^4\rhob_{4\itxt} \\[5pt]
& \pbar_\| && :=\Ppar && +\eps\,G_{1\itxt}^\| && +\eps^2 G_{2\itxt}^\| && +\eps^3 G_{3\itxt}^\| \\[5pt]
& \upmubar && :=\upmuhat && +\eps\,G_{1\itxt}^\upmu && +\eps^2 G_{2\itxt}^\upmu && +\eps^3 G_{3\itxt}^\upmu \\[5pt]
& \thbar && :=\Theta && +\eps\,G_{1\itxt}^\Theta && +\eps^2 G_{2\itxt}^\Theta && +\eps^3 G_{3\itxt}^\Theta
\end{alignedat}$ \\
\hline
$\begin{gathered}
\text{gyrocenter coordinates} \\[2pt] \Zbcal:=(\Xb,\Ppar,\mu,\Theta)
\end{gathered}$ &
$\mu:=\upmuhat+\eps\gavg{\gamma_2^\Theta}+\eps^2\gavg{\gamma_3^\Theta}$ \\
\hline
\end{tabu}
\end{center}
\caption{Coordinate changes for ions involved in the phase-space coordinate transformation
$(\Xb,\Ppar,\mu,\Theta)\mapsto(\xb,\vb)$ relating physical coordinates and gyrocenter coordinates.}
\label{tab_super_ions}
\end{table}
\begin{table}
\begin{center}
\tabulinesep=2mm
\begin{tabu}{|c|c|}
\hline
$\begin{gathered}
\text{physical coordinates} \\ (\xb,\vb)
\end{gathered}$ & \\
\hline 
$\begin{gathered}
\text{Hamiltonian picture} \\ (\xb,\pb)
\end{gathered}$ &
$\pb:=\vb-\sqrt{\eps}\,\Ab_1$ \\
\hline
$\begin{gathered}
\text{angle coordinates} \\ (\xb,\ppar,\upmu,\theta)
\end{gathered}$ &
$\begin{alignedat}{2}
& \ppar  && :=\pb\cdot\bb_0 \\[1mm]
& \upmu  && :=\frac{1}{2B_0}\left|\bb_0\times\pb\times\bb_0\right|^2 \\
& \theta && :=\arctan\left(\frac{\pb\cdot\eb_1}{\pb\cdot\eb_2}\right)
\end{alignedat}$ \\[-1mm]
\hline
$\begin{gathered}
\text{guiding-center coordinates} \\ \Zbbar:=(\Xbbar,\pbar_\|,\upmubar,\thbar)
\end{gathered}$ &
$\begin{alignedat}{5}
& \xb && :=\Xbbar && +\eps\,\bar\rhob_{1\etxt} && +\eps^2\bar\rhob_{2\etxt} &&
+\eps^3\bar\rhob_{3\etxt}+\eps^4\bar\rhob_{4\etxt} \\[5pt]
& \ppar && :=\pbar_\| && +\eps\,\Gbar_{1\etxt}^\| && +\eps^2\Gbar_{2\etxt}^\| && +\eps^3\Gbar_{3\etxt}^\| \\[5pt]
& \upmu && :=\upmubar && +\eps\,\Gbar_{1\etxt}^\upmu && +\eps^2\Gbar_{2\etxt}^\upmu && +\eps^3\Gbar_{3\etxt}^\upmu \\[5pt]
& \theta && :=\thbar && +\eps\,\Gbar_{1\etxt}^\Theta && +\eps^2\Gbar_{2\etxt}^\Theta && +\eps^3\Gbar_{3\etxt}^\Theta
\end{alignedat}$ \\
\hline
$\begin{gathered}
\text{preliminary gyrocenter coordinates} \\ \Zb:=(\Xb,\Ppar,\upmuhat,\Theta)
\end{gathered}$ &
$\begin{alignedat}{4}
& \Xbbar   && :=\Xb && +\eps^2\rhob_{2\etxt} && +\eps^\fivehalves\rhob_{\fivehalves\etxt}
+\eps^3\rhob_{3\etxt} \,, \\[5pt]
& \pbar_\| && :=\Ppar    && +\eps\,G_{1\etxt}^\|     && +\eps^2 G_{2\etxt}^\| \,, \\[5pt]
& \upmubar && :=\upmuhat && +\sqrt{\eps}\,G_{\onehalf\etxt}^\upmu  && +\eps\,G_{1\etxt}^\upmu \,, \\[5pt]
& \thbar   && :=\Theta   && +\sqrt{\eps}\,G_{\onehalf\etxt}^\Theta && +\eps\,G_{1\etxt}^\Theta \,.
\end{alignedat}$ \\
\hline
$\begin{gathered}
\text{gyrocenter coordinates} \\ \Zbcal:=(\Xb,\Ppar,\mu,\Theta)
\end{gathered}$ &
$\mu:=\upmuhat+\sqrt{\eps}\gavg{G_{\onehalf\etxt}^\upmu}-\eps\gavg{\gamma_3^\Theta}$ \\
\hline
\end{tabu}
\end{center}
\caption{Coordinate changes for electrons involved in the phase-space coordinate transformation
$(\Xb,\Ppar,\mu,\Theta)\mapsto(\xb,\vb)$ relating physical coordinates and gyrocenter coordinates.}
\label{tab_super_electrons}
\end{table}
The first transformation moves the magnetic vector potential $\Ab_1$ from the symplectic
part of the single-particle Lagrangian to the Hamiltonian, by defining the ``momentum''
\be
\pb:=
\begin{dcases}
\vb+\eps\,\Ab_1 & \text{ions} \,, \\
\vb-\sqrt{\eps}\Ab_1 & \text{electrons} \,.
\end{dcases}
\ee
This is a near-identity transformation in $\vb$ in the limit $\eps \to 0$, with unit
Jacobian determinant. It resembles the usual transformation to canonical coordinates,
but it does not contain the background magnetic vector potential $\Ab_0$. The new
single-particle Lagrangians read
\begin{subequations}
\begin{align}
 L_\itxt &=  \bigg(\pb + \frac{\Ab_0}{\eps} \bigg)
\cdot\ldot{\xb} - \left( \frac{|\pb|^2}{2} + \eps\, \Psi_{1\itxt} + \eps^2 \frac{|\Ab_1|^2}{2} \right)  
 \,, 
 \\[1mm]
 L_\etxt &=  \bigg(\sqrt{\eps}\,\pb - \frac{\Ab_0}{\eps} \bigg)
\cdot\ldot{\xb} - \left( \frac{|\pb|^2}{2} - \sqrt{\eps}\,\Psi_{1\etxt} + \eps\, \frac{|\Ab_1|^2}{2} 
\right) \,, 
\end{align}
\end{subequations}
where we introduced the \emph{generalized potentials}
\be
\Psi_{1\stxt}:=
\begin{dcases}
\phi_1-\pb\cdot\Ab_1 & \text{ions} \,, \\
\sqrt{\eps}\,\phi_1-\pb\cdot\Ab_1 & \text{electrons} \,.
\end{dcases}
\ee
This first preliminary transformation is not necessary for performing the gyrokinetic
reduction. However, it leads to simpler calculations in the following. Since all
dynamic potentials now occur in the Hamiltonian, the $(\xb,\pb)$-coordinate representation
is also called the \emph{Hamiltonian picture}. We also remark that electromagnetic
gauge invariance has been broken by this preliminary coordinate transformation.
We then introduce local cylindrical coordinates in $\pb$-space, namely
\be \label{transf:v}
p_\parallel:=\pb\cdot\bb_0 \,, \qquad
\upmu:=\frac{|\bb_0\times\pb\times\bb_0|^2}{2 B_0} \,, \qquad
\theta:=\arctan\bigg(\frac{\pb\cdot\eb_1}{\pb\cdot\eb_2}\bigg) \,,
\ee
where $(\eb_1,\eb_2,\bb_0)$ represents a local static orthonormal basis of $\RR^3$,
given an arbitrary unit vector $\eb_1$ perpendicular to $\bb_0$. Denoting by $\pb_\perp
:=\bb_0\times\pb\times\bb_0$ the component of $\pb$ perpendicular to the local background
magnetic field, we have $\pb=\ppar\bb_0+\pb_\perp$, with $\pb_\perp=(\pb\cdot\eb_1)\eb_1
+(\pb\cdot\eb_2)\eb_2$. From the definition of $\theta$ we can write $(\pb\cdot\eb_1)
=-\sqrt{2\upmu B_0}\sin\theta$ and $(\pb\cdot\eb_2)=-\sqrt{2\upmu B_0}\cos\theta$
and thus define a second $\theta$-dependent orthonormal basis $(\ab_0,\bb_0,\cb_0)$, with
$\ab_0:=\eb_1\cos\theta-\eb_2\sin\theta$ and $\cb_0:=-\eb_1\sin\theta-\eb_2\cos\theta$.
We note that $\bb_0\times\cb_0=\ab_0$, $\de\ab_0/\de\theta=\cb_0$ and $\de\cb_0/\de\theta=-\ab_0$,
which will be used in later calculations. The transformation to angle coordinates thus reads
\be \label{tau:prime}
 \pb=\ppar\bb_0 + \sqrt{2 \upmu B_0}\,\cb_0 \,,
\ee
with Jacobian determinant $B_0$. This leads to the single-particle Lagrangians
\begin{subequations}
\begin{align}
 L_\itxt &=  \bigg(\ppar\bb_0 + \sqrt{2 \upmu B_0}\, \cb_0 + \frac{\Ab_0}{\eps} \bigg)
\cdot\ldot{\xb} - \left( \frac{\ppar^2}{2}+\upmu B_0 + \eps\, \Psi_{1\itxt} + \eps^2\, 
\frac{|\Ab_1|^2}{2} \right) \,,
\label{Li:start}
 \\[1mm]
 L_\etxt &=  \bigg(\sqrt{\eps}\,\ppar\bb_0 + \sqrt{\eps}\sqrt{2 \upmu B_0}\, \cb_0 - 
\frac{\Ab_0}{\eps} \bigg) \cdot\ldot{\xb} - \left( \frac{\ppar^2}{2}+\upmu B_0 
-\sqrt{\eps}\, \Psi_{1\etxt} + \eps\, 
\frac{|\Ab_1|^2}{2} \right) \,.
\label{Le:start}
\end{align}
\end{subequations}

\subsection{Guiding-center coordinates $(\Xbbar,\pbar_\para,\upmubar,\thbar)$}
The guiding-center phase-space coordinate transformation dates back to the pioneering
work of \citep{Littlejohn1983} and has the purpose of removing the gyro-angle dependence
from those parts of the Lagrangians \eqref{Li:start}-\eqref{Le:start} that do not
depend on the fluctuating potentials $\phi_1$ and $\Ab_1$.
For this reason, the gyrokinetic
literature often describes the guiding-center phase-space coordinate transformation
as a transformation acting on a single-particle Lagrangian that involves only quantities
related to the background magnetic field $\Bb_0$ and does not feature any fluctuating
fields. These are said to be added at a later stage, after the guiding-center coordinate
transformation has been performed. We believe that this description is slightly misleading,
as it seems to suggest the idea that the single-particle Lagrangian is modified by adding
terms related to the fluctuating fields during the process of transforming the phase-space
coordinates. In fact, the fluctuating fields are present in the single-particle Lagrangian
since the beginning of the derivation (as it should be, once we identify the physical system
that we want to describe), but their gyro-angle dependence is simply treated at a later
stage, after the guiding-center coordinate transformation has been performed. Following
\citep{Possanner2018}, we define the guiding-center coordinate transformation as
a polynomial transform of the form
\begin{subequations}
\label{poly:transf:gc}
\begin{alignat}{5}
& \xb    && :=\Xbbar      && +\eps\,\rhobbar_{1\stxt}     && +\eps^2\rhobbar_{2\stxt}     && 
+\eps^3\rhobbar_{3\stxt}+\eps^4\rhobbar_{4\stxt} \,,
 \\[1mm]
& \ppar  && :=\pbar_\para && +\eps\,\Gbar_{1\stxt}^\para  && +\eps^2\Gbar_{2\stxt}^\para  && 
+\eps^3\Gbar_{3\stxt}^\para\,, 
\\[1mm]
& \upmu  && :=\upmubar    && +\eps\,\Gbar_{1\stxt}^\upmu  && +\eps^2\Gbar_{2\stxt}^\upmu  && 
+\eps^3\Gbar_{3\stxt}^\upmu 
\,,
\\[1mm]
& \theta && :=\thbar      && +\eps\,\Gbar_{1\stxt}^\Theta && +\eps^2\Gbar_{2\stxt}^\Theta && 
+\eps^3\Gbar_{3\stxt}^\Theta\,,
\end{alignat}
\end{subequations}
where $\rhobbar_{n\stxt}$, $\Gbar^\para_{n\stxt}$, $\Gbar^\upmu_{n\stxt}$ and $\Gbar^\Theta_{n\stxt}$
denote the generators of the coordinate transformation for the respective particle species.
The generators are functions of the \emph{guiding-center coordinates}
$\Zbbar:=(\Xbbar,\pbar_\para,\upmubar,\thbar)$
and may additionally depend on time. The guiding-center coordinates are the guiding-center
position $\Xbbar$, the guiding-center parallel momentum $\pbar_\para$, the guiding-center
magnetic moment $\upmubar$ and the guiding-center angle variable $\thbar$. The idea
is then to substitute the coordinate transformation \eqref{poly:transf:gc} in the
single-particle Lagrangians \eqref{Li:start}-\eqref{Le:start}, by using the transformation
law of vector fields
\be \label{tangentmap}
 \ldot \xb = \ldot \Xbbar + \sum_{n=1}^{4} \eps^n\, \ldot \rhobbar_{n\stxt}
=\ldot \Xbbar + \sum_{n=1}^{4} \eps^n\,\left(
\pder{\rhobbar_{n\stxt}}{\Zbbar} \cdot \ldot \Zbbar + \pder{\rhobbar_{n\stxt}}{t}\right) \,,
\ee
to obtain the corresponding guiding-center Lagrangians. The work of \citep{Possanner2018}
showed that the gyro-angle dependence due to the term $\sqrt{2\upmu B_0}\,\cb_0$
can be indeed removed via polynomial transforms in maximal ordering. We repeat these
calculations in Appendix \ref{app:gc} and arrive at
\begin{subequations}
\begin{align}
 L_\itxt &\sim \left(\pbar_\para\bb_0+\frac{\Ab_0}{\eps}\right)\cdot\ldot{\Xbbar}
 + \eps\,\upmubar\,\ldot{\thbar} - \Big[ \Hbar_0 + \eps\,\Hbar_{1\itxt} + \eps^2 \Hbar_{2\itxt} 
 + \overline O(\eps^3)\Big] + 
O(\eps^{4})\,, \label{Li:gc} \\[1mm]
\begin{split}
 L_\etxt &\sim  \left(\sqrt{\eps}\,\pbar_\para\bb_0 - \frac{\Ab_0}{\eps}\right)\cdot\ldot{\Xbbar}
 - \eps^2\,\upmubar\,\ldot{\thbar} \\[1mm]
 &\quad - \left[ \Hbar_0 + \sqrt \eps\,\Hbar_{\onehalf\etxt} + \eps\, \Hbar_{1\etxt} + 
\eps^{\threehalves} \Hbar_{\threehalves\etxt} + \overline 
O(\eps^{2}) \right] + O(\eps^{4}) \,,
\end{split} \label{Le:gc}
\end{align}
\end{subequations}
where the symbol $\sim$ denotes the equivalence between Lagrangians, namely the fact
that two Lagrangians differ only by the total differential of some scalar function.
Moreover, the symbol $\Obar(\eps^n)$ denotes corrections to the Hamiltonians
of order $O(\eps^n)$ that are independent of the guiding-center angle $\thbar$. The
ion and electron guiding-center Hamiltonians in \eqref{Li:gc}-\eqref{Le:gc} read
\begin{subequations}
\label{Hgc}
\begin{align}
 \Hbar_0 & :=  \frac{\pbar_\para^2}{2}+\upmubar B_0 \,, && \Hbar_0 :=  \frac{\pbar_\para^2}{2} 
+ \upmubar B_0 \,,
 \\[2mm]
 \Hbar_{1\itxt} & :=  \Psi_\itxt + \delta H_{1} \,,  && \Hbar_{\onehalf\etxt} := \pb \cdot \Ab_1\,,
 \\[1mm]
 \Hbar_{2\itxt} & :=  \onehalf|\Ab_1|^2 + \delta H_{2}\,, && \Hbar_{1\etxt} := -\phi_1 + 
\onehalf|\Ab_1|^2\,,
 \\[1mm]
 & && \Hbar_{\threehalves\etxt} := -\delta H_1\,.
\end{align}
\end{subequations}

We remark the following comments about the guiding-center single-particle Lagrangians
\eqref{Li:gc}-\eqref{Le:gc}: \\

\begin{itemize}
 \item The dynamic potentials in the Hamiltonians \eqref{Hgc} are evaluated at the
physical particle position $\xb/\eps$:
\begin{subequations}
\label{phi_in_arg}
\begin{align}
\phi_1 \left( t, \frac{\xb}{\eps} \right) & = \phi_1 \left( t, \frac{\Xbbar}{\eps} + 
\rhobbar_{1\stxt} + \eps\, \rhobbar_{2\stxt} + O(\eps^2) \right)\,,
 \\[1mm]
\Ab_1 \left( t, \frac{\xb}{\eps} \right) & = \Ab_1 \left( t, \frac{\Xbbar}{\eps} + 
\rhobbar_{1\stxt} + \eps\, \rhobbar_{2\stxt} + O(\eps^2) \right)\,.
\end{align}
\end{subequations}
The gyro-angle dependence in the generators $\rhobbar_{n\stxt}$ occurring in the
arguments of the fluctuating potentials will be removed eventually by the
transformation from guiding-center to gyrocenter coordinates, as discussed
in detail in the next section. \\

\item Due to our assumption of maximal ordering, the guiding-center Hamiltonians
feature geometric terms related to the curvature of the background magnetic field,
in particular
\be
\label{delH1i}
\delta H_{1}=\upmubar\left[
\frac{\pbar_\para}{2}\left(\Nabla\times\bb_0\right)\cdot\bb_0
-\pbar_\para \left(\Nabla\ab_0\cdot\cb_0\right)\cdot\bb_0\right] \,,
\ee
and a cumbersome term $\delta H_2$, which can be deduced from \eqref{def:dH2} in
the appendix. The two terms in \eqref{delH1i} are usually referred to as Ba\~nos
drift \citep{Banos1967} and gyro-gauge term, respectively. The curvature terms are
less important for the electrons, where $\delta H_1$ appears at order $O(\eps^\threehalves)$,
because of the mass ratio between ions and electrons of order $O(\sqrt\eps)$. \\

\item For electrons, the magnetic perturbations $\Ab_1$ are $O(\sqrt{\eps})$ larger
than the electric perturbations $\phi_1$. This can be already foreseen in the normalized
Vlasov-Maxwell equations \eqref{LT} and is due to the mass ratio between ions and
electrons. Moreover, it shows the importance of electron dynamics in electromagnetic
gyrokinetic simulations of fusion plasmas. \\

\item Due to the error term $\Obar(\eps^2)$ in the electron Hamiltonian, the electron
guiding-center single-particle Lagrangian \eqref{Le:gc}
is less accurate than the ion guiding-center single-particle Lagrangian \eqref{Li:gc}.
This is due to the fact that the guiding-center magnetic moment $\upmubar$ has been
computed with less precision for electrons than for ions.
We could easily
improve the accuracy of $\upmubar$ for electrons, but, as we can see from \eqref{Le:gc},
the dynamic potentials $\phi_1$ and $\Ab_1$ play a more prominent role than any curvature
terms. In the Hamiltonian, the term $\pb\cdot\Ab_1$ of $\Psi_1$ appears at order $O(\sqrt\eps)$ and the term
$|\Ab_1|^2/2$ appears at order $O(\eps)$, whereas the first curvature terms appear at
$O(\eps^\threehalves)$. This is in contrast to the ions, where the first curvature term $\delta H_1$
appears already at order $O(\eps)$, which is the same order as $\Psi_1$ and one order lower
than the quadratic term $|\Ab_1|^2/2$. In order to achieve an equally accurate description
for the electrons, we should truncate the electron single-particle Lagrangian at
order $O(\eps^5)$: this is beyond the scope of the work presented here,
but does not represent a limitation of the method in general. \\

\item The Jacobian determinants $J_\stxt$ of the guiding-center transformation
$\Zbbar \mapsto (\xb,\vb)$ can be computed directly from the symplectic part of
the guiding-center single-particle Lagrangians \eqref{Li:gc}-\eqref{Le:gc}:
\begin{subequations}
\label{jac_gc}
\begin{align}
 J_\itxt &= B^*_{\para \itxt} = B_0
+ \eps\,\pbar_\para \left(\Nabla \times \bb_0\right)\cdot\bb_0 \,,
 \\[1mm]
 J_\etxt &= B^*_{\para\etxt} = B_0
- \eps^\threehalves\,\pbar_\para \left(\Nabla \times \bb_0\right)\cdot\bb_0 \,. 
\end{align}
\end{subequations}
Such Jacobian determinants are exact because the symplectic forms in \eqref{Li:gc}-\eqref{Le:gc}
remain the same at any order of the guiding-center expansion, as only the guiding-center
Hamiltonians change with increased order of accuracy (see, for example, \citep{Possanner2018}
for a proof of this statement). The Jacobian determinants confirm that geometric terms
related to the curvature of the background magnetic field appear at order $O(\eps)$ for the
ions and at order $O(\eps^\threehalves)$ for the electrons, in accordance with the guiding-center
Hamiltonians \eqref{Hgc}.
\end{itemize}

\subsection{Gyrocenter coordinates $(\Xb,P_\para,\mu,\Theta)$} \label{transf:gy}

The guiding-center single-particle Lagrangians \eqref{Li:gc}-\eqref{Le:gc} obtained
from the guiding-center coordinate transformation still carry a dependence on the
guiding-center angle $\thbar$ (the fast variable) in the arguments of the dynamic
potentials \eqref{phi_in_arg}. Consequently, the guiding-center magnetic moment
$\upmubar$ is not a constant of the motion and the dynamics of slow and fast variables
are still coupled in the guiding-center phase space. The purpose of the gyrocenter
phase-space coordinate transformation is to remove this residual dependence on the
angle variable $\thbar$ from the Lagrangians, thus from the Hamiltonians
\eqref{Hgc}. As for the guiding-center coordinate transformation \eqref{poly:transf:gc},
we define the gyrocenter coordinate transformation for \emph{ions} as polynomial transforms
of the form
\begin{subequations}
\label{poly:transf:gy}
\begin{alignat}{4}
\Xbbar & =\Xb && +\eps^2\rhob_{2\itxt} && +\eps^3\rhob_{3\itxt} && +\eps^4\rhob_{4\itxt} \,, \\
\pbar_\para & =\Ppar && +\eps\, G_{1\itxt}^\para && +\eps^2 G_{2\itxt}^\para &&
+\eps^3 G_{3\itxt}^\para \,, \\[1mm]
\upmubar & =\upmuhat && +\eps\, G_{1\itxt}^\upmu && +\eps^2 G_{2\itxt}^\upmu &&
+\eps^3 G_{3\itxt}^\upmu \,, \\[1mm]
\thbar & =\Theta && +\eps\, G_{1\itxt}^\Theta && +\eps^2 G_{2\itxt}^\Theta &&
+\eps^3 G_{3\itxt}^\Theta \,,
\end{alignat}
\end{subequations}
and the gyrocenter coordinate transformation for \emph{electrons} as polynomial transforms
of the form
\begin{subequations}
\label{poly:transf:gy_el}
\begin{alignat}{2}
\Xbbar & = \Xb && + \eps^2 \rhob_{2\etxt} + \eps^{\fivehalves}\rhob_{\fivehalves\etxt}
+ \eps^3 \rhob_{3\etxt} \,, \\[0.5mm]
\pbar_\para & = P_\para && +\eps\,G_{1\etxt}^\para +\eps^2 G_{2\etxt}^\para \,, \\[1.5mm]
\upmubar & = \upmuhat && +\sqrt\eps\, G_{\onehalf\etxt}^\upmu +\eps\, G_{1\etxt}^\upmu \,, \\[1mm]
\thbar & =\Theta && +\sqrt\eps\, G_{\onehalf\etxt}^\Theta + \eps\, G_{1\etxt}^\Theta \,.  
\end{alignat}
\end{subequations}
Here, $\Zb:=(\Xb,\Ppar,\upmuhat,\Theta)$ denote preliminary gyrocenter coordinates,
and $\rhob_{n\stxt}$, $G_{n\stxt}^\para$, $G_{n\stxt}^\upmu$ and $G_{n\stxt}^\Theta$
(with $n$ integer or half-integer) denote the generators of the coordinate transformation
for the respective particle species. Our preliminary gyrocenter coordinates are the
gyrocenter position $\Xb$, the gyrocenter parallel momentum $\Ppar$, the preliminary
gyrocenter magnetic moment $\upmuhat$ and the gyrocenter angle variable $\Theta$,
also called \mbox{\emph{gyro-angle}}. The polynomial transform for the electrons
is defined by polynomials
in powers of $\sqrt\eps$ because of the mass ratio between ions and electrons. Moreover,
it consists of fewer terms than the ion coordinate transformation because of the lower
accuracy of the electron guiding-center single-particle Lagrangian. If more accuracy
is desired, the number of terms in the polynomial transform can be increased, but
this goes beyond the scope of this work. We also set $\rhob_{1\stxt}=0$ a priori:
it is, in principle, possible to keep these first-order generators in the calculations
and then find out that they can be set to zero without loss of generality.

We remark the conceptual simplicity of the polynomial transform $\Zb\mapsto\Zbbar$
compared to Lie transforms \citep{Brizard2007}: for each coordinate, the transformation
is a polynomial of finite degree in $\eps$ (the degree being adapted to the desired
accuracy of the transformation) and it is moreover linear and algebraic in the generators.
By substituting \eqref{poly:transf:gy}-\eqref{poly:transf:gy_el} in the Lagrangians
\eqref{Li:gc}-\eqref{Le:gc},
the gyrocenter generators can be chosen in order to eliminate the residual dependence
on the gyro-angle $\Theta$. The method is analogous to the guiding-center transformation
and it is discussed in detail in \citep{Possanner2018} for the long-wavelength regime,
that is, the case of dynamic potentials with spatial variations on the macroscopic
length scale $\xhat$. In what follows we apply the same methodology to the short-wavelength
(strongly-varying) regime expressed in \eqref{phi_in_arg}.

The exact same ideas and computations of the guiding-center transformation can be
applied also for the gyrocenter transformation. In particular, we make use of the
equivalence of Lagrangians under the addition of the total differential $\ldot S$
of arbitrary scalar functions $S(t,\Zb)$ and write
\be
L_\itxt \sim L_\itxt + \eps^2 \ldot S_{2\itxt} + \eps^3 \ldot S_{3\itxt} \,, \qquad
L_\etxt \sim L_\etxt + \eps^{\fivehalves} \ldot S_{\fivehalves\etxt} + \eps^3 \ldot S_{3\etxt} \,,
\ee
where the total differential $\ldot S_{n\stxt}$ reads
\be \label{def:dotS}
 \ldot S_{n\stxt} = \frac{1}{\eps}\Nabla_\perp S_{n\stxt} \cdot \ldot \Xb + \nabla_\para 
S_{n\stxt}\,\bb_0 \cdot \ldot \Xb 
+ \pder{S_{n\stxt}}{\Ppar}\ldot \Ppar + \pder{S_{n\stxt}}{\upmuhat} \ldot \upmuhat + 
\pder{S_n\stxt}{\Theta} \ldot \Theta + \pder{S_{n\stxt}}{t} \,.
\ee
Here, $\Nabla_\perp:= -\bb_0 \times \Nabla \times \bb_0$ and $\nabla_\para:= \bb_0 \cdot \nabla$
denote the gradients with respect to the direction perpendicular and parallel to
the background magnetic field, respectively. We remark that in the derivation of
the gyrokinetic Lagrangians, the scalar functions $S_{n\stxt}$ turn out to be functions
of the fluctuating potentials $\phi_1$ and $\Ab_1$ and thus have strong variations
in the perpendicular directions, which has to be expressed in \eqref{def:dotS} by
means of the factor $1/\eps$ in front of $\Nabla_\perp$. \\

We summarize our results for ions in Propositions 1-3 and our results for electrons
in Propositions 4-6. Proofs of these propositions are given in Appendix \ref{app_proofs}.
The species index is mostly omitted for more readability. \\

{\bf Proposition 1 (ion polynomial transform) }
{\it The ion guiding-center single-particle Lagrangian
\eqref{Li:gc}, expressed in the preliminary gyrocenter coordinates $(\Xb,\Ppar,\upmuhat,\Theta)$
via the polynomial transform \eqref{poly:transf:gy}, is equivalent to
\be
\label{prop:i:Li}
L_\itxt \sim \left( P_\para\bb_0 + \frac{\Ab_0}{\eps} \right) \cdot \ldot \Xb - H_0 + \sum_{n=1}^3 
\eps^n L_n + O(\eps^4)\,,
\ee
where $H_0 = P_\para^2/2 + \upmuhat\, B_0$ is the lowest-order Hamiltonian and the Lagrangians 
$L_n$ read
\be \label{gy:Ln_prop}
L_n = \gab_n^{\Xb} \cdot \ldot \Xb + \gamma_n^\para\, \ldot P_\para + \gamma_n^{\upmu}\, \ldot 
\upmuhat + \gamma_n^\Theta\, \ldot \Theta - H_n \,,
\ee
where the components $\gab_n^{\Xb},\gamma_n^\para,\gamma_n^{\upmu},\gamma_n^\Theta$
and the Hamiltonians $H_n$ depend on the generators of the transformation \eqref{poly:transf:gy}
and are given in Appendix \ref{app_proofs}.
} \\

{\bf Proposition 2 (preliminary ion gyrocenter Lagrangian)}
{\it In the Lagrangians $L_n$ of \eqref{gy:Ln_prop}
the generators of the polynomial transform \eqref{poly:transf:gy} can be chosen such that
\be \label{Li:prelim}
L_\itxt \sim \left(\Ppar\bb_0+\frac{\Ab_0}{\eps}\right)\cdot\ldot\Xb
+\eps\left(\upmuhat+\eps\,\gavg{\gamma_2^\Theta}+\eps^2\gavg{\gamma_3^\Theta}\right)\ldot\Theta
- H_0 + O(\eps^4) \,,
\ee
where $\gavg{\gamma_2^\Theta}$ and $\gavg{\gamma_3^\Theta}$ are given in \eqref{gam2:gavg}
and \eqref{gam3:res} in Appendix \ref{app_proofs} and, for a given function $g(\Theta)$, $\gavg g$ denotes
its gyro-average and is defined as
\begin{equation}
\label{def:gavg}
\gavg g :=\frac{1}{2\pi}\int_0^{2\pi}\dd\Theta\,g(\Theta) \,.
\end{equation}
} \\

{\bf Proposition 3 (ion gyrocenter Lagrangian)}
{\it The generalized gyrocenter magnetic moment
\be \label{def:mu}
\mu:=\upmuhat+\eps \gavg{\gamma_2^\Theta}+\eps^2 \gavg{\gamma_3^\Theta}
\ee
is a constant of the motion, accurate up to order $O(\eps^2)$, with respect to the
dynamics induced by the preliminary ion gyrocenter single-particle Lagrangian \eqref{Li:prelim}.
Moreover, there is a one-to-one correspondence $\mu \mapsto \upmuhat$, which implies
that 
\be
\Hcal_\itxt:=\frac{\Ppar^2}{2}+\upmuhat(\mu)B_0
\ee
is the ion gyrocenter Hamiltonian.
In other words, $\Hcal_\itxt$ is obtained from $H_0$ by inverting the
transformation $\upmuhat \mapsto \mu$ defined in \eqref{def:mu}. By expressing
\eqref{Li:prelim} in terms of the new gyrocenter coordinates $\Zbcal:=(\Xb,\Ppar,\mu,\Theta)$,
we obtain the ion gyrocenter single-particle Lagrangian
\be
\label{ion_gyrocenter_lag}
\Lcal_\itxt\sim\left(\Ppar\bb_0+\frac{\Ab_0}{\eps}\right)\cdot\ldot\Xb+\eps\mu\,\ldot\Theta
- \left[\Hcal_\itxt + \Obar(\eps^3)\right] + O(\eps^4)  \,,
\ee
where the symbol $\Obar(\eps^3)$ denotes corrections to the Hamiltonian of order $O(\eps^3)$
that are independent of the gyro-angle $\Theta$ and $\Hcal_\itxt=\Hcal_{0\itxt}
+\eps\,\Hcal_{1\itxt}+\eps^2\,\Hcal_{2\itxt}$, with
\begin{subequations}
\label{ion_gyrocenter_ham}
\begin{align}
\Hcal_{0\itxt}= &\: \frac{\Ppar^2}{2}+\mu B_0 \,, \\[2mm]
\Hcal_{1\itxt}= &\: \gavg{\Psi_1}+\delta H_1 \,, \\[1mm]
\begin{split}
\Hcal_{2\itxt}= &\: \frac{1}{2}\gavg{|\Ab_1|^2}-\frac{1}{2 B_0}\der{}{\mu}\gavg{\widetilde{\Psi_1}^2} \\
& -\frac{1}{2B_0^2}\gavg{\left(\bb_0\times\nabla_\perp\widetilde{\Psi_1}\right)\cdot \Nabla_\perp
\int^\Theta\dd\Theta'\,\widetilde{\Psi_1}} \\
& -\frac{1}{B_0}\gavg{\nabla_\para \wt{\Psi_1}\int^\Theta\dd\Theta'\,\widetilde{A_{1\|}}}
+ \delta\Gcal_2 + \delta H_2 \,.
\end{split}
\end{align}
\end{subequations}
Here,
\be
\Psi_1=\Psi_1\left(t,\frac{\Xb}{\eps}+\rhobbar_{1\itxt}\right) \quad \text{with}
\quad \rhobbar_{1\itxt}=\sqrt{\frac{2\mu}{B_0(\Xb)}}\ab_0(\Xb,\Theta) \,,
\ee
and $\wt{\Psi_1}:=\Psi_1-\gavg{\Psi_1}$ (and the same for $\Ab_1$). Moreover,
$\delta\Gcal_2$ is a term related to the curvature of the background magnetic field, besides
$\delta H_1$ and $\delta H_2$:
\be \label{def:dG2star}
\begin{aligned}
\delta\Gcal_2:=\delta G_2 &- \frac 12 \gavg{
\frac{\widetilde{\Psi_1}}{B_0}
\left(\frac{\bb_0}{B_0}\times \Nabla_\perp B_0\right) \cdot \Nabla_\perp
\int^\Theta\dd\Theta'\,\frac{\widetilde{\Psi_1}}{B_0}} \\[1mm]
& + \frac 12 \gavg{\left(\frac{\bb_0}{B_0^3}\times \Nabla_\perp\widetilde{\Psi_1}\right)
\cdot \Nabla_\perp B_0 \int^\Theta\dd\Theta'\,\widetilde{\Psi_1}} \,,
\end{aligned}
\ee
with $\delta G_2$ given in \eqref{def:delG2} in Appendix \ref{app_proofs}.
} \\

{\bf Proposition 4 (electron polynomial transform) }
{\it The electron guiding-center single-particle
Lagrangian \eqref{Le:gc}, expressed in the preliminary gyrocenter coordinates $(\Xb,\Ppar,\upmuhat,\Theta)$
via the polynomial transform \eqref{poly:transf:gy_el}, is equivalent to
\be \label{prop:e:Le}
L_\etxt \sim \left( \sqrt\eps P_\para\bb_0 - \frac{\Ab_0}{\eps} \right) \cdot \ldot \Xb - H_0 - 
\sqrt\eps\, H_{\onehalf} + \sum_{n=2}^6 \eps^{\frac{n}{2}} L_{\frac{n}{2}} + O(\eps^\sevenhalves)\,,
\ee
where $H_0= P_\para^2/2 + \upmuhat\,B_0$ is the lowest-order Hamiltonian, the Hamiltonian
$H_\onehalf$ reads
\be
 H_\onehalf = G_\onehalf^\upmu B_0 + P_\para A_{1\para} + \sqrt{2\upmuhat B_0}\,\cb_0\cdot 
\Ab_{1\perp} \,, \label{def:Hhalb}
\ee
and the Lagrangians $L_{\frac{n}{2}}$ read
\be \label{e:gy:Ln_prop}
L_{\frac{n}{2}} = \gab_{\frac{n}{2}}^{\Xb} \cdot \ldot \Xb + \gamma_{\frac{n}{2}}^\para\, \ldot P_\para
+ \gamma_{\frac{n}{2}}^{\upmu}\, \ldot \upmuhat + \gamma_{\frac{n}{2}}^\Theta\,\ldot\Theta-H_{\frac{n}{2}} \,,
\ee
where the components $\gab_{\frac{n}{2}}^{\Xb},\gamma_{\frac{n}{2}}^\para,\gamma_{\frac{n}{2}}^{\upmu},
\gamma_{\frac{n}{2}}^\Theta$ and the Hamiltonians $H_{\frac{n}{2}}$ depend on the
generators of the transformation \eqref{poly:transf:gy_el} and are given in Appendix \ref{app_proofs}.
} \\

{\bf Proposition 5 (preliminary electron gyrocenter Lagrangian)}
{\it In the Hamiltonian $H_\onehalf$ and in the Lagrangians $L_{\frac{n}{2}}$ of
\eqref{e:gy:Ln_prop} the generators of the polynomial transform \eqref{poly:transf:gy_el}
can be chosen such that
\be \label{Le:prelim}
\begin{aligned}
L_\etxt &\sim \left( \sqrt\eps\, P_\para\bb_0 - \frac{\Ab_0}{\eps} \right) \cdot \ldot \Xb
- \eps^2 \left(\upmuhat + \sqrt\eps\,\gavg{G_\onehalf^\upmu} - \eps\gavg{\gamma_3^\Theta} \right)\ldot\Theta
- H_0 + O(\eps^\sevenhalves) \,,
 \end{aligned}
\ee
where $\gavg{G_\onehalf^\upmu}$ and $\gavg{\gamma_3^\Theta}$ are given in \eqref{e:gavgG1}
and \eqref{gam3th_el_final} in Appendix \ref{app_proofs} and, for a given function
$g(\Theta)$, $\gavg g$ denotes its gyro-average as defined in \eqref{def:gavg}.
} \\

{\bf Proposition 6 (electron gyrocenter Lagrangian)}
{\it The generalized gyrocenter magnetic moment
\be \label{e:def:mu}
\mu:=\upmuhat + \sqrt\eps\,\gavg{G_\onehalf^\upmu} - \eps\gavg{\gamma_3^\Theta}
\ee
is a constant of the motion, accurate up to order $O(\eps)$, with respect to the
dynamics induced by the preliminary electron gyrocenter single-particle Lagrangian
\eqref{Le:prelim}. Moreover, there is a one-to-one correspondence $\mu\mapsto\upmuhat$,
which implies that
\be
\Hcal_\etxt:=\frac{\Ppar^2}{2}+\upmuhat(\mu)B_0
\ee
is the electron gyrocenter Hamiltonian. In other words, $\Hcal_\etxt$ is obtained
from $H_0$ by inverting the
transformation $\upmuhat \mapsto \mu$ defined in \eqref{e:def:mu}. By expressing
\eqref{Le:prelim} in terms of the new gyrocenter coordinates $\Zbcal:=(\Xb,\Ppar,\mu,\Theta)$,
we obtain the electron gyrocenter single-particle Lagrangian
\be
\label{electron_gyrocenter_lag}
\Lcal_\etxt\sim\left(\sqrt\eps\,\Ppar\bb_0 - \frac{\Ab_0}{\eps}\right)\cdot\ldot\Xb - 
\eps^2\mu\,\ldot\Theta
- \left[\Hcal_\etxt + \Obar(\eps^\threehalves)\right] + O(\eps^\sevenhalves) \,,
\ee
where the symbol $\Obar(\eps^\threehalves)$ denotes corrections to the Hamiltonian
of order $O(\eps^\threehalves)$ that are independent of the gyro-angle $\Theta$ and
$\Hcal_\etxt=\Hcal_{0\etxt}+\sqrt\eps\,\Hcal_{\onehalf\etxt}+\eps\,\Hcal_{1\etxt}$, with
\begin{subequations}
\label{electron_gyrocenter_ham}
\begin{align}
\Hcal_{0\etxt}= &\: \frac{\Ppar^2}{2}+\mu B_0 \,, \\[2mm]
\Hcal_{\onehalf\etxt}= &\: P_\para A_{1\para} \,, \\[1mm] 
\Hcal_{1\etxt}= &\: - \phi_1 + \frac{A_{1\para}^2}{2} + \mu\,(\Nabla\times\Ab_1)\cdot\bb_0 \,.
\end{align}
\end{subequations}
Here, $\phi_1=\phi_1(t,\Xb/\eps)$ and the same for $\Ab_1$.
}

\section{Gyrokinetic Vlasov-Maxwell model}
We first remark that the Jacobian determinants $\Jcal_\stxt$ of the gyrocenter coordinate
transformation $\Zbcal\mapsto(\xb,\vb)$ can be computed directly from the symplectic
part of the gyrocenter single-particle Lagrangians \eqref{ion_gyrocenter_lag} and
\eqref{electron_gyrocenter_lag} for ions and electrons, respectively:
\begin{subequations}
\label{jacobians_gyro}
\begin{align}
\label{jacobian_gyro_ion}
\Jcal_\itxt & =\Bstar_{\|\itxt} =B_0+\eps\,\Ppar\left(\Nabla\times\bb_0\right)\cdot\bb_0 \,, \\
\label{jacobian_gyro_el}
\Jcal_\etxt & =\Bstar_{\|\etxt}=B_0-\eps^\threehalves\Ppar\left(\Nabla\times\bb_0\right)\cdot\bb_0 \,.
\end{align}
\end{subequations}
Such Jacobian determinants are exact and have the same form as the corresponding
guiding-center Jacobian determinants in \eqref{jac_gc} because the symplectic forms
in \eqref{ion_gyrocenter_lag} and \eqref{electron_gyrocenter_lag} remain the same
at any order of the gyrocenter expansion, as only the gyrocenter Hamiltonians change
with increased order of accuracy (as for the guiding-center coordinate transformation).
The Jacobian determinants
\eqref{jacobians_gyro} confirm that geometric terms related to the curvature of
the background magnetic field appear again at order $O(\eps)$ for the ions and at
order $O(\eps^\threehalves)$ for the electrons.

The ion gyrokinetic equations of motion for the slow phase-space variables $(\Xb,\Ppar)$
derived from the ion gyrocenter Lagrangian~\eqref{ion_gyrocenter_lag} read
\begin{subequations}
\label{ion_eom}
\begin{align}
\ldot\Xb    & =\frac{1}{\Bstar_{\|\itxt}}\left(\eps\,\bb_0\times\Nabla\Hcal_\itxt
+\frac{\de\Hcal_\itxt}{\de\Ppar}\Bbstar_\itxt\right)+O(\eps^3) \,, \\[5pt]
\ldot\Ppar  & =-\frac{\Bbstar_\itxt}{\Bstar_{\|\itxt}}\cdot\Nabla\Hcal_\itxt+O(\eps^3) \,, 
\end{align}
\end{subequations}
where the modified magnetic field $\Bbstar_\itxt$ is defined as $\Bbstar_\itxt:=
\Bb_0+\eps\,\Ppar\,\Nabla\times\bb_0$ and its parallel component is defined as
$\Bstar_{\|\itxt}:=\Bbstar_\itxt\cdot\bb_0$, which is the ion Jacobian \eqref{jacobian_gyro_ion}.
The gyrokinetic magnetic moment $\mu$ is a constant of the motion accurate up to
order $O(\eps^3)$: $\ldot\mu=O(\eps^3)$. Moreover, the dynamics of the gyro-angle
$\Theta$ is decoupled from the slow dynamics of $(\Xb,\Ppar)$ and described by
$\ldot\Theta=(1/\eps)\de\Hcal_\itxt/\de\mu+O(\eps^2)$, with the factor $1/\eps$
signifying that this dynamics is the fastest among all phase-space variables and
with larger error terms of order $O(\eps^2)$. The ion gyrocenter distribution function
$F_\itxt(t,\Xb,\Ppar,\mu)$ is constant along solutions of \eqref{ion_eom}, where
$\mu$ is a time-independent parameter.
The electron gyrokinetic equations of motion for the slow phase-space variables $(\Xb,\Ppar)$
derived from the electron gyrocenter Lagrangian \eqref{electron_gyrocenter_lag} read
\begin{subequations}
\label{electron_eom}
\begin{align}
\ldot\Xb    & =-\frac{1}{\sqrt{\eps}}\frac{1}{\Bstar_{\|\etxt}}\left(\eps^\threehalves\,
\bb_0\times\Nabla\Hcal_\etxt-\frac{\de\Hcal_\etxt}{\de\Ppar}\Bbstar_\etxt\right)+O(\eps) \,, \\[5pt]
\ldot\Ppar  & =-\frac{1}{\sqrt{\eps}}\frac{\Bbstar_\etxt}{\Bstar_{\|\etxt}}\cdot\Nabla\Hcal_\etxt
+O(\eps) \,, 
\end{align}
\end{subequations}
where the modified magnetic field $\Bbstar_\etxt$ is defined as $\Bbstar_\etxt:=
\Bb_0-\eps^\threehalves\,\Ppar\,\Nabla\times\bb_0$ and its parallel component is
defined again as $\Bstar_{\|\etxt}:=\Bbstar_\etxt\cdot\bb_0$, which is the electron
Jacobian~\eqref{jacobian_gyro_el}. The gyrokinetic magnetic moment $\mu$ is a constant
of the motion accurate up to order~$O(\eps^\threehalves)$: $\ldot\mu=O(\eps^\threehalves)$.
Moreover, the dynamics of the gyro-angle $\Theta$ is decoupled from the slow dynamics
of $(\Xb,\Ppar)$ and described by $\ldot\Theta=-(1/\eps^2)\de\Hcal_\etxt/\de\mu
+O(\eps^{-\onehalf})$, with the factor $1/\eps^2$ signifying again that this dynamics
is the fastest among all phase-space variables and with larger error terms of order
$O(\eps^{-\onehalf})$. The electron gyrocenter distribution function
$F_\etxt(t,\Xb,\Ppar,\mu)$ is constant along solutions of \eqref{electron_eom}, where
$\mu$ is a time-independent parameter.

The non-homogeneous gyrokinetic Maxwell's equations can be derived from the variational
principle by taking variations of the Low Lagrangian with respect to the electromagnetic
fluctuating potentials $\Phi_1$ and $\Ab_1$. After applying identity \eqref{ess:equal},
the weak form of gyrokinetic Coulomb's
law is obtained by taking variations with respect to $\Phi_1$ and reads
\begin{equation*}
\begin{split}
0= &\: \int\dd^6\Zbcal\bigg(\Bstarpare F_\etxt\gavg{\delta\Phi_1}
-\Bstarpari F_\itxt\gavg{\delta\Phi_1}\bigg) \\[5pt]
& +\eps\int\dd^6\Zbcal\,\Bstarpari F_\itxt\Bigg(\frac{1}{B_0}\der{}{\mu}
\gavg{\widetilde{\Psi_{1\itxt}}\,\widetilde{\delta\Phi_1}} \\[5pt]
& +\frac{1}{2B_0^2}\gavg{\left(\bb_0\times\nabla_\perp\widetilde{\delta\Phi_1}\right)\cdot
\int^\Theta\dd\Theta'\,\nabla_\perp\widetilde{\Psi_{1\itxt}}} \\[5pt]
& +\frac{1}{2B_0^2}\gavg{\left(\bb_0\times\nabla_\perp\widetilde{\Psi_{1\itxt}}\right)\cdot
\int^\Theta\dd\Theta'\,\nabla_\perp\widetilde{\delta\Phi_1}} \\[5pt]
& +\frac{1}{B_0}\gavg{\nabla_\|\delta\Phi_1\int^\Theta\dd\Theta'\,\widetilde{A_{1\|}}}
+\frac{\delta(\delta\Gcal_2)}{\delta\Phi_1}(\delta\Phi_1)\Bigg) \,,
\end{split}
\end{equation*}
where $\delta\Phi_1$ denotes an arbitrary test function. Here we neglected all terms
of order higher than $O(\eps^2)$ and $O(\eps)$ from the ion and electron Hamiltonians,
respectively. The terms of order $O(1)$ and $O(\eps)$ represent the gyrocenter
charge density and the gyrocenter polarization density, respectively. Similarly,
after applying again identity \eqref{ess:equal}, the weak form of Amp\`{e}re-Maxwell's law is obtained by taking variations with respect to $\Ab_1$ and reads
\begin{equation*}
\begin{split}
0 = &
-\int\dd^3\xb\,\bigg(\nabla\times\Bb_0+\eps\,\nabla\times(\nabla\times\Ab_1)\bigg)\cdot\delta\Ab_1 \\[5pt]
&-\sqrt{\eps}\int\dd^6\Zbcal\,\Bstarpare F_\etxt\,\Ppar\,\delta A_{1\|} 
+\eps\int\dd^6\Zbcal\Bigg[\Bstarpari F_\itxt\gavg{\Pb\cdot\delta\Ab_1} \\[5pt]
& -\Bstarpare F_\etxt\bigg(A_{1\|}\delta A_{1\|}+\mu\,(\Nabla\times\delta\Ab_1)\cdot\bb_0\bigg)\Bigg] \\[5pt]
&-\eps^2\int\dd^6\Zbcal\,\Bstarpari F_\itxt\Bigg(\frac{1}{B_0}\der{}{\mu}
\gavg{\widetilde{\Psi_{1\itxt}}\,\Pb\cdot\widetilde{\delta\Ab_1}}+\gavg{\Ab_1\cdot\delta\Ab_1} \\[5pt]
& +\frac{1}{2B_0^2}\gavg{\left(\bb_0\times\nabla_\perp(\Pb\cdot\widetilde{\delta\Ab_1})\right)\cdot
\int^\Theta\dd\Theta'\,\nabla_\perp\widetilde{\Psi_{1\itxt}}} \\[5pt]
& +\frac{1}{2B_0^2}\gavg{\left(\bb_0\times\nabla_\perp\widetilde{\Psi_{1\itxt}}\right)\cdot
\int^\Theta\dd\Theta'\,\nabla_\perp(\Pb\cdot\widetilde{\delta\Ab_1})} \\[5pt]
& +\frac{1}{B_0}\gavg{\nabla_\|(\Pb\cdot\delta\Ab_1)\int^\Theta\dd\Theta'\,\widetilde{A_{1\|}}}
+\frac{1}{B_0}\gavg{\nabla_\|\Psi_{1\itxt}\int^\Theta\dd\Theta'\,\widetilde{\delta A_{1\|}}} \\[5pt]
& +\frac{\delta(\delta\Gcal_2)}{\delta\Ab_1}(\delta\Ab_1)\Bigg) \,,
\end{split}
\end{equation*}
where $\delta\Ab_1$ denotes an arbitrary test function and ${\Pb:=\Ppar\bb_0
+\sqrt{2\mu B_0}\,\cb_0}$.
As before, we neglected
all terms of order higher than $O(\eps^2)$ and $O(\eps)$ from the ion and electron
Hamiltonians, respectively. \\

\section{Conclusions and outlook}
The main results of this work are summarized in Proposition 3 for ions and in Proposition
6 for electrons. We state the final ion and electron single-particle gyrocenter Lagrangians
\eqref{ion_gyrocenter_lag} and \eqref{electron_gyrocenter_lag} in normalized form,
obtained from the consecutive coordinate transformations listed in Tables~\ref{tab_super_ions}
and \ref{tab_super_electrons}, respectively.
The results obtained in \citep{Tronko2018} and in the previous works of \citep{Hahm1988}
and \citep{Brizard1989} for ions are recovered and augmented by terms related to
the assumption of maximal ordering. In particular, novel terms are the geometric
first-order and second-order corrections $\delta H_1$, $\delta H_2$ and $\delta\Gcal_2$,
appearing in the first-order and second-order ion gyrocenter Hamiltonians $\Hcal_{1\itxt}$
and $\Hcal_{2\itxt}$ in \eqref{ion_gyrocenter_ham}, respectively. In particular,
the first-order curvature term
\be
\delta H_{1}=\mu\left[
\frac{\Ppar}{2}\left(\Nabla\times\bb_0\right)\cdot\bb_0
-\Ppar\left(\Nabla\ab_0\cdot\cb_0\right)\cdot\bb_0\right]
\ee
should not be neglected when computing ion trajectories in maximal ordering, since
it is of the same size as the curvature term in $\Bbstar_\itxt$.
Moreover, the term
\begin{equation}
-\frac{1}{B_0}\gavg{\nablapar\wt{\Psi_1}\int^\Theta\dd\Theta'\,\wt{A_{1\|}}} \,,
\end{equation}
appearing in the second-order ion gyrocenter Hamiltonian $\Hcal_{2\itxt}$, is also new.
It is important to note that this term arises under the conventional gyrokinetic ordering
$\kpar/|\kb_\perp|=O(\eps)$ (see discussions in section \ref{sec:phys}) and it is related to
fluctuations with non-zero toroidal mode number.

Concerning the electrons, they turn out to be insensitive to magnetic background 
curvature effects up to second order in $\eps$, even
in maximal ordering. However, terms related to the parallel component of the fluctuating magnetic
vector potential appear in the electron gyrocenter Hamiltonian already at order $O(\sqrt{\eps})$.
Moreover, the first-order term $\mu(\Nabla\times\Ab_1)\cdot\bb_0=\mu B_{1\|}$ represents
a correction to the perpendicular electron kinetic energy, which now reads $\mu(B_0+\eps B_{1\|})$.
This correction is absent for the ions at first order.

The ion gyrokinetic equations of motion \eqref{ion_eom} for the slow variables $(\Xb,\Ppar)$
are accurate up to order $O(\eps^3)$. The same accuracy is achieved for the conservation
of the gyrocenter magnetic moment $\mu$: the corrections $\Obar(\eps^3)$ to the
ion gyrocenter Hamiltonian in \eqref{ion_gyrocenter_lag} are indeed independent of $\Theta$
and do not play a role in the
Euler-Lagrange equation $\dd/\dd t(\de\Lcal_\itxt/\de\ldot\Theta)=\de\Lcal_\itxt/\de\Theta$.

The electron gyrokinetic equations of motion \eqref{electron_eom} for the slow
variables $(\Xb,\Ppar)$ are accurate up to order $O(\eps)$. By contrast,
the conservation of the gyrocenter magnetic moment $\mu$ is accurate up to
order $O(\eps^\threehalves)$: the corrections $\Obar(\eps^\threehalves)$ to the
electron gyrocenter Hamiltonian in \eqref{electron_gyrocenter_lag} are indeed independent
of $\Theta$ and do not play a role
in the Euler-Lagrange equation $\dd/\dd t(\de\Lcal_\etxt/\de\ldot\Theta)=\de\Lcal_\etxt/\de\Theta$).

The electron gyrokinetic Lagrangian, and the corresponding gyrokinetic equations
of motion, have been derived within an ordering consistent with the ions, despite
the order of accuracy of the results being different for the two species (due to
the fact that the gyrocenter magnetic moment $\mu$ has been computed with less precision
for electrons than for ions). We conclude that it is possible to derive a set of
gyrokinetic Vlasov-Maxwell equations for ions and electrons within our unique methodology
based on guiding-center and gyrocenter polynomial transforms and within unique ordering
assumptions relevant for realistic fusion scenarios (maximal ordering). Our technique
is alternative to the use of Lie transforms and, combined with our rigorous normalization
procedure, can provide useful insights into the derivation of gyrokinetic models
and a solid starting point for their further rigorous mathematical investigation.

\vspace{2em}
We would like to thank Eric Sonnendr\"ucker for supporting our research work and
Bruce Scott, Roman Hatzky and Cesare Tronci for insightful discussions about many of the topics treated
in this article. We would like to thank also Yaman G\"u\c cl\"u for useful suggestions
about how to present some of the material discussed here as clearly as possible for
the non-specialist reader.
This work has been carried out within the framework of the EUROfusion
Consortium and has received funding from the Euratom research and training program
2014-2018 and 2019-2020 under grant agreement No 633053. The views and opinions
expressed herein do not necessarily reflect those of the European Commission.

\appendix

\section{Variational principle}
\label{app_variat}
In \eqref{def:Lp} the single-particle Lagrangian $L_\stxt$ is evaluated at the $t$-family
of maps ${\Psi_t: \mathbb R^6 \to \mathbb R^6}$, parametrically depending on time.
The map $\Psi_t$ is the flow map of the characteristics of the Vlasov equation \eqref{Vlasov},
\be
\label{chars}
\der{\xb}{t} = \vb \,, \qquad 
\der{\vb}{t}=\frac{q_\stxt}{m_\stxt}\Big[\Eb(t,\xb)+\vb\times\Bb(t,\xb)\Big] \,. 
\ee
The flow transports a particle that is at $\zb_0 = (\xb_0,\vb_0)$ at the initial time $t_0$ to the phase space 
position $\Psi_t(\zb_0)$ at time $t$; it is a volume-preserving diffeomorphism, $\Psi_t 
\in 
\tn{Diff}_\tn{vol}(\mathbb R^6)$. The description of an ensemble of particles via the particle 
Lagrangian $L_\tn{p}$ arises from the single picture in the following way: 
Newton's equation of motion for a charged particle can be deduced from the variational 
principle
\be \label{delta:z}
 \delta \int_{t_0}^{t_1} L_\tn{s} \left( \zb(t),\dt{\zb(t)} \right)\,\tn d t = 0\,.
\ee
The extremum defined by \eqref{delta:z} is denoted by $\zb(t) = (\xb(t),\vb(t))$ and is the 
solution of the Euler-Lagrange equations, given by \eqref{chars}. Hence, $\zb(t) = 
\Psi_t(\zb_0)$ and \eqref{delta:z} can be written as
\be \label{delta:psi}
 \delta \int_{t_0}^{t_1} \int \delta^6(\zb_1 - \zb_0)\, L_\tn{s} \left( 
\Psi_t(\zb_1),\dt{\Psi_t(\zb_1)} 
\right) \,\tn d^6 \zb_1\, \tn d t = 0\,.
\ee
Formally, the ensemble description is obtained by replacing the delta function with the initial 
particle distribution, $\delta^6(\zb_1 - \zb_0) \to f_{\stxt,0}(\zb_1)$, which yields
\be \label{delta:psi:f}
 \delta \int_{t_0}^{t_1} \int f_{\stxt,0}(\zb_1)\, L_\tn{s} \left( 
\Psi_t(\zb_1),\dt{\Psi_t(\zb_1)} 
\right) \,\tn d^6 \zb_1\, \tn d t = 0\,,
\ee
The particle Lagrangian $L_\tn{p}$ from \eqref{def:Lp} is obtained by taking the sum over the 
species 
and by relabeling the variables of integration $\zb_1 \to \zb_0$. By construction, the variation 
\eqref{delta:psi:f} with respect to $\Psi_t$ yields the characteristics of the Vlasov 
equation. The 
Vlasov equation itself enters the picture via the definition 
\be \label{def:f}
 f_\stxt(t,\zb) := \int \delta^6(\zb - \Psi_t(\zb_0))\, f_{\stxt,0}(\zb_0)\,\tn d^6\zb_0
\ee
of the particle distribution functions.
Remark that $\zb \in \mathbb R^6$ denote coordinates here and not a path in phase space. Hence 
\eqref{def:f} makes the link between Lagrangian paths $\Psi_t$ and the distribution function 
$f_\stxt$ via $f_\stxt(t,\zb) = f_{\stxt,0}((\Psi_t)^{-1}(\zb))$, or simply $f_\stxt = 
f_{\stxt,0} 
\circ (\Psi_t)^{-1}$. This implies in particular that $f_s$ is constant along the Lagrangian 
paths,
\be
 f_\stxt(t,\Psi_t(\zb_0)) = f_{\stxt,0}(\zb_0)\,,
\ee
which is the statement of the Vlasov equation. Let us now write the characteristic equations in the 
form
\be
\dt{\zb}=\Fb(\zb) \,, \qquad \zb(t=0)=\zb_0 \,,
\ee
with the vector field $\Fb=(\vb,\ab)$, where $\ab=(q_\stxt/m_\stxt)(\Eb+\vb\times\Bb)$.
From the definition of the distribution function it follows that
\be
\label{ess:equal}
 \int\dd^6\zb_0\,f_{\stxt,0}(\zb_0)\,
L_\tn{s}\left(\Psi_t,\dt{\Psi_t}\right) = 
\int\dd^6\zb\,f_\stxt(t,\zb)\,L_\stxt(\zb,\Fb(\zb)) \,.
\ee
This is easily verified by substituting the definition \eqref{def:f} into the right-hand 
side and preforming the integration over $\zb$.

\section{Guiding-center transformation}
\label{app:gc}

Here we review how the ion and electron guiding-center Lagrangians \eqref{Li:gc}-\eqref{Le:gc}
are obtained from the polynomial transforms \eqref{poly:transf:gc}, following
\citep{Possanner2018}. In order to shorten the calculations, we use the coordinate
$p_\perp:=\sqrt{2\upmu B_0}$ instead of the magnetic moment $\upmu$ introduced in
\eqref{transf:v}. Hence, we have the momentum transformation
\be
\label{def:pperp}
p_\perp:=\pbar_\perp+\eps\,\Gbar^\perp_{1\stxt}+\eps^2\,\Gbar^\perp_{2\stxt}
+\eps^3\,\Gbar^\perp_{3\stxt} \,,
\ee
and denote by $\Zbbar_\perp:=(\Xbbar,\pbar_\para,\pbar_\perp,\thbar)$ the guiding-center
coordinates. The final result is then transformed back to the representation in terms of
the magnetic moment. Let us start with the ions and consider the starting Lagrangian 
\eqref{Li:start}. We substitute the polynomial transform \eqref{poly:transf:gc},
with $p_\perp$ from \eqref{def:pperp} instead of $\upmu$, and expand in Taylor series
around $\Zbbar$. For instance,
\be \label{A01}
\begin{aligned}
 \Ab_0(\xb) &= \Ab_0(\Xbbar + \eps\,\rhobbar_{1\itxt} + 
\eps^2\rhobbar_{2\itxt} + \eps^3\rhobbar_{3\itxt} + \eps^4\rhobbar_{4\itxt})
 \\[2mm]
 &= \Ab_0(\Xbbar) + \left[\left( \eps\,\rhobbar_{1\itxt} + 
\eps^2\rhobbar_{2\itxt} + \eps^3\rhobbar_{3\itxt} + \eps^4\rhobbar_{4\itxt} \right) \cdot \Nabla 
\right] \Ab_0(\Xbbar)
 \\[1mm]
 &\qquad\qquad\: \, + \frac{1}{2} \left[\left( \eps\,\rhobbar_{1\itxt} + 
\eps^2\rhobbar_{2\itxt} + \eps^3\rhobbar_{3\itxt} + \eps^4\rhobbar_{4\itxt} \right) \cdot \Nabla 
\right]^2 \Ab_0(\Xbbar) +\: \dots
\end{aligned}
\ee
We also use the equivalence of Lagrangians under the addition of total differentials
of arbitrary scalar functions and the transformation law of tangents \eqref{tangentmap}
(here applied to $\dot \Ab_0(\Xbbar)$) to write 
\be 
\begin{split}
\Ab_0(\Xbbar) \cdot \ldot \rhobbar_{n\itxt}(\Xbbar,\pbar_\para,\pbar_\perp,\thbar,t) & \sim 
-\ldot\Ab_0(\Xbbar) \cdot \rhobbar_{n\itxt}(\Xbbar,\pbar_\para,\pbar_\perp,\thbar,t) \\[1mm]
& =-\ldot\Xbbar\cdot\Nabla\Ab_0(\Xbbar) \cdot 
\rhobbar_{n\itxt}(\Xbbar,\pbar_\para,\pbar_\perp,\thbar,t) \,.
\end{split}
\ee
We then encounter terms like
\be \label{A03}
\begin{aligned}
 \left[- \Nabla \Ab_0(\Xbbar) \cdot 
\rhobbar_{n\itxt} + \rhobbar_{n\itxt} \cdot \Nabla \Ab_0(\Xbbar) \right] \cdot \ldot \Xbbar &= 
- \left[ \rhb_{n\itxt} \times (\Nabla \times \Ab_0(\Xbbar)) \right] \cdot \ldot \Xbbar
 \\[1mm]
 &= - \left[ \rhb_{n\itxt} \times \Bb_0(\Xbbar)\right] \cdot \ldot \Xbbar\,,
\end{aligned}
\ee
in the transformed Lagrangian. The computations for polynomial transforms of arbitrary
order have been carried out in \citep{Possanner2018}. We repeat here in particular the
results from Proposition~1 on page~12 of this work to write the series expansion
of the Lagrangian, up to order $N=3$ (omitting the species index for more readability).
The new Lagrangian written without the dynamical potentials $\phi_1$ and $\Ab_1$,
which are not transformed in the guiding-center step, reads
\be \label{Li:expand}
 L_\itxt \sim \frac{\Lbar_{-1}}{\eps} + \Lbar_0 + \eps\, \Lbar_1 + \eps^2 
\Lbar_2 + \eps^3 \Lbar_3 + O(\eps^4)\,,
\ee
where $\Lbar_{-1} = \Ab_0 \cdot \ldot \Xbbar$ and
\begin{align}
 \Lbar_0 &= (\pbar_\para \bb_0 + \pbar_\perp\cb_0 - \rhobbar_1 \times \Bb_0) \cdot \ldot 
\Xbbar -  \frac{\pbar_\para^2}{2} - \frac{\pbar_\perp^2}{2}\,,
\label{L0}
 \\[1mm]
 \Lbar_1 &= (\Gbar_1^\para \bb_0 + \Gbar_1^\perp \cb_0 - \rhobbar_2 \times \Bb_0 + \Qbcal_1) 
\cdot \ldot \Xbbar - 
 \pbar_\para \Gbar_1^\para - \pbar_\perp \Gbar_1^\perp  + \Lbcal_1\,,  
\label{L1}
 \\[1mm]
 \Lbar_2 &= (\Gbar_2^\para \bb_0 + \Gbar_2^\perp \cb_0 - \rhobbar_3 \times \Bb_0 + \Qbcal_2) 
\cdot \ldot \Xbbar - 
 \pbar_\para \Gbar_2^\para - \pbar_\perp \Gbar_2^\perp + \Lbcal_2\,,  
\label{L2}
 \\[1mm]
 \Lbar_3 &= (\Gbar_3^\para \bb_0 + \Gbar_3^\perp \cb_0 - \rhobbar_4 \times \Bb_0 + \Qbcal_3) 
\cdot \ldot \Xbbar - 
 \pbar_\para \Gbar_3^\para - \pbar_\perp \Gbar_3^\perp + \Lbcal_3\,.  
\label{L3}
\end{align}
In order to determine the guiding-center Lagrangians we need the explicit expressions of 
\begin{align}
 \Qbcal_1 &:= \frac{1}{2}(\rhobbar_1 \cdot \Nabla \Bb_0) \times \rhobbar_1 - \pbar_\para \rhobbar_1 
\times (\Nabla \times \bb_0) - \pbar_\perp \rhobbar_1 \times (\Nabla \times \cb_0) - \pbar_\perp 
\Gbar_1^\Theta \ab_0\,,
\end{align}
and of
\begin{align*}
 \Lbcal_1 := & - \frac{1}{2} \ldot \rhobbar_1 \cdot (\rhobbar_1 \times \Bb_0) - \ldot \pbar_\para 
(\rhobbar_1 \cdot \bb_0) - \ldot \pbar_\perp (\rhobbar_1 \cdot \cb_0) + \ldot 
\thbar\,\pbar_\perp(\rhobbar_1 \cdot \ab_0)\,,
\\[2mm]
\begin{split}
 \Lbcal_2 := & - \ldot \rhobbar_1 \cdot (\rhobbar_2 \times \Bb_0) - \ldot \Gbar_1^\para 
(\rhobbar_1 \cdot \bb_0) - \ldot \Gbar_1^\perp (\rhobbar_1 \cdot \cb_0) + 
 \ldot \thbar\,\Gbar_1^\perp (\rhobbar_1 \cdot \ab_0)
 \\[2mm]
 & - \ldot \pbar_\para (\rhobbar_2 \cdot \bb_0) - \ldot \pbar_\perp (\rhobbar_2 \cdot \cb_0) + 
 \ldot \thbar\,\pbar_\perp(\rhobbar_2 \cdot \ab_0)   \nonumber
 \\[2mm]
 & + \pbar_\para\, \rhobbar_1 \cdot \Nabla \bb_0 \cdot \ldot \rhobbar_1 + \pbar_\perp\, 
\rhobbar_1 \cdot \Nabla \cb_0 \cdot \ldot \rhobbar_1 - \pbar_\perp \Gbar_1^\Theta\,\ldot \rhobbar_1 
\cdot \ab_0   \nonumber
 \\[1mm]
 & - \frac{1}{3} \ldot \rhobbar_1 \cdot [\rhobbar_1 \times (\rhobbar_1 \cdot \Nabla) \Bb_0] - 
\frac{1}{6} \rhobbar_1 \times [(\rhobbar_1 \cdot \Nabla)^2 \Bb_0] \cdot \ldot \Xbbar\,,
\end{split}
\end{align*}
\begin{align*}
\begin{split}
 \Lbcal_3 := & - \frac{1}{2} \ldot \rhobbar_2 \cdot (\rhobbar_2 \times \Bb_0) - \ldot \rhobbar_1 \cdot 
(\rhobbar_3 \times \Bb_0)
 \\[1mm]
 & - \ldot \Gbar_2^\para (\rhobbar_1 \cdot \bb_0) - \ldot \Gbar_2^\perp
(\rhobbar_1 \cdot \cb_0) + \ldot \thbar\, \Gbar_2^\perp (\rhobbar_1 \cdot \ab_0)  \nonumber
 \\[1mm]
 & - \ldot \Gbar_1^\para (\rhobbar_2 \cdot \bb_0) - \ldot \Gbar_1^\perp (\rhobbar_2 \cdot \cb_0) 
 + \ldot \thbar\,\Gbar_1^\perp (\rhobbar_2 \cdot \ab_0)   \nonumber
 \\[2mm]
 & - \ldot \pbar_\para (\rhobbar_3 \cdot \bb_0) - \ldot \pbar_\perp (\rhobbar_3 \cdot \cb_0) + 
 \ldot \thbar\,\pbar_\perp(\rhobbar_3 \cdot \ab_0)   \nonumber
 \\[1mm]
 & + \sum_{m=1}^2 \Gbar_{2-m}^\para\, \rhobbar_1 \cdot \Nabla \bb_0 \cdot \ldot \rhobbar_m + 
\pbar_\para\, \left[\rhobbar_2 \cdot \Nabla \bb_0 + \frac{1}{2}(\rhobbar_1 \cdot \Nabla)^2 \bb_0 
\right] \cdot \ldot \rhobbar_1 \nonumber
 \\[0mm]
 & + \sum_{m=1}^2 \Gbar_{2-m}^\perp \left[ \rhobbar_1 \cdot \Nabla \cb_0 - \Gbar_1^\Theta 
\ab_0 \right]\cdot \ldot \rhobbar_m   \nonumber
 \\[0mm]
 & + \pbar_\perp\, \bigg[\rhobbar_2 \cdot \Nabla \cb_0 - \Gbar_2^\Theta\,\ab_0 + 
\frac{1}{2} \bigg(\rhobbar_1 \cdot \Nabla + \Gbar_1^\Theta \pder{}{\thbar} \bigg)^2 \cb_0 
\bigg] \cdot \ldot \rhobbar_1   \nonumber
 \\[0mm]
 & - \ldot \rhobbar_1 \cdot [\rhobbar_2 \times (\rhobbar_1 \cdot \Nabla) \Bb_0] - \frac{3}{24} 
\ldot \rhobbar_1 \cdot [\rhobbar_1 \times (\rhobbar_1 \cdot \Nabla)^2 \Bb_0]  \nonumber
 \\[1mm]
 & - \frac{1}{2} \rhobbar_2 \times [(\rhobbar_1 \cdot \Nabla)^2 \Bb_0] \cdot \ldot \Xbbar 
 - \frac{1}{24} \rhobbar_1 \times [(\rhobbar_1 \cdot \Nabla)^3 \Bb_0] \cdot \ldot \Xbbar\,,
\end{split}
\end{align*}
We remark that the above expansions are straightforward but cumbersome, in particular
at higher orders, as in $\Lbcal_3$ for example. One could automatize these expansions
in a symbolic computer program, similar to the ideas in \citep{Burby}, where a different
approach of setting up the guiding-center transform has been implemented.

In the following we choose the generators $\rhobbar_n,\Gbar_n^\para,\Gbar_n^\perp$
and $\Gbar_n^\Theta$ in order to cancel as many terms as possible from the Lagrangians.
In $\Lbar_0$ from \eqref{L0} we require $\pbar_\perp\cb_0 = \rhobbar_1 \times \Bb_0$,
which can be obtained by setting
\be \label{rhob_i}
\rhobbar_1 = \frac{\pbar_\perp}{B_0}\,\ab_0\,.
\ee
Moreover, it has been shown in \citep[Theorem 1]{Possanner2018} that the remaining 
generators in \eqref{L1}-\eqref{L3} can be chosen such that 
\be \label{L123i}
 \Lbar_1 \sim \frac{\pbar_\perp^2}{2 B_0} \,\ldot \thbar\,,\qquad \Lbar_2 \sim 
-\frac{\delta H_1}{B_0}\,\ldot 
\thbar\,,\qquad \Lbar_3 \sim -\frac{\delta H_2}{B_0}\,\ldot \thbar\,,
\ee
where $-\delta H_1/B_0$ and $-\delta H_2/B_0$ are the $\thbar$-averages of the terms 
multiplying $\ldot \thbar$ in $\Lbcal_2$ and $\Lbcal_3$ of \eqref{L2} and \eqref{L3}, respectively. It is mandatory to keep these 
terms in the Lagrangian to avoid secularities in the averaged equations of motion. Substituting 
\eqref{rhob_i} and \eqref{L123i} into 
\eqref{Li:expand} yields 
\be \label{Li:gc:new}
 L_\itxt \sim \left(\pbar_\para \bb_0 + \frac{\Ab_0}{\eps}\right) \cdot \ldot 
\Xbbar + \eps\left( \frac{\pbar_\perp^2}{2 B_0} - \eps\,\frac{\delta H_1}{B_0} - 
\eps^2\,\frac{\delta H_2}{B_0} \right)\ldot \thbar - 
 \frac{\pbar_\para^2}{2} - \frac{\pbar_\perp^2}{2} + O(\eps^4)\,,
\ee
without the dynamical potentials $\phi_1$ and $\Ab_1$. 
Computing 
the Euler-Lagrange equation $\pa L_\itxt/\pa \thbar - \dt{} \pa L_\itxt/\pa \dot \thbar = 0$ and 
noting that \eqref{Li:gc:new} is independent of $\thbar$ up to third order, we obtain
\be \label{dtmu:i}
 \dt{} \left( \frac{\pbar_\perp^2}{2 B_0} - \eps\, \frac{\delta H_1}{B_0} - \eps^2 \frac{\delta 
H_2}{B_0} \right) = O(\eps^3)\,.
\ee
Hence, the guiding-center \emph{generalized magnetic moment}
\be \label{def:mubar}
 \upmubar := \frac{\pbar_\perp^2}{2 B_0} - \eps\, \frac{\delta H_1}{B_0} - \eps^2 
\frac{\delta H_2}{B_0}
\ee
is a constant of the motion accurate up to order $O(\eps^3)$ and should be adopted as one of 
the coordinates. As indicated by the notation $\delta H_1$ and $\delta H_2$, there is a 
one-to-one correspondence between the guiding-center generalized magnetic moment
$\upmubar$ and the guiding-center Hamiltonian $\Hbar$:
\be
 \Hbar := \frac{\pbar_\para^2}{2} + \frac{\pbar_\perp^2(\upmubar)}{2}\,.
\ee
In other words, the guiding-center Hamiltonian is obtained by expressing $\pbar_\perp^2/2$
in terms of $\upmubar$ by inverting the transformation $\pbar_\perp \mapsto \upmubar$
defined in \eqref{def:mubar}. This one-to-one correspondence is typical for polynomial 
transforms and occurs also in the gyrocenter transformation. It has an important
consequence for the accuracy of the derived guiding-center Lagrangian, namely it
yields
\be \label{Li:gc:mu}
 L_\itxt \sim \left(\pbar_\para \bb_0 + \frac{\Ab_0}{\eps}\right) \cdot \ldot 
\Xbbar + \eps\,\upmubar\,\ldot \thbar - \left[\Hbar + \Obar(\eps^3) \right] + 
O(\eps^4)\,.
\ee
Here, the notation $\Obar(\eps^3)$ denotes corrections to the Hamiltonian $\Hbar$ that 
are of order $O(\eps^3)$ and independent of $\thbar$, originating from the inversion of 
\eqref{def:mubar}.
The loss of one order of accuracy in the Hamiltonian from \eqref{Li:gc:new} to \eqref{Li:gc:mu} 
occurs because the generalized magnetic moment $\upmubar$ has been determined only
up to $O(\eps^2)$ in \eqref{def:mubar}, implying that the guiding-center Hamiltonian 
$\Hbar$ can only be second-order accurate. Is is important to note that neglecting the 
second-order term $\delta H_2$ in \eqref{Li:gc:new} leads to a guiding-center Hamiltonian
that is only first-order accurate. We arrive to a similar conclusion for the electrons below.

Let us now identify the higher-order terms $\delta H_1$ and $\delta H_2$ in the
guiding-center generalized magnetic moment \eqref{def:mubar}. As mentioned earlier,
they are the $\thbar$-averages of the terms multiplying $\ldot \thbar$ in \eqref{L2}
and \eqref{L3}, respectively. From \eqref{L2} and \eqref{L3} we observe that all
$\ldot\thbar$-terms are in $\Lbcal_2$ and $\Lbcal_3$. From \eqref{rhob_i} we deduce that
\be
\rhobbar_1 \cdot \bb_0 = \rhobbar_1 \cdot \cb_0 = 0\,,\qquad \ldot \rhobbar_1 = 
\pder{\rhobbar_1}{\thbar}\,\ldot \thbar + \tn{rest} = \frac{\pbar_\perp}{B_0}\cb_0\,\ldot \thbar + 
\tn{rest}\,.
\ee
Therefore, the term multiplying $\ldot \thbar$ in $\Lbcal_2$, denoted by 
$\Lbcal_2^\theta$, reads
\be \label{L2:terms}
\begin{aligned}
 \Lbcal_2 &\: =\Lbcal_2^\theta\,\ldot\thbar + \tn{rest}
 \\[2mm]
 &\:=  \left[ - \frac{\pbar_\perp}{B_0}\cb_0 \cdot (\rhobbar_2 \times \Bb_0) + 
\Gbar_1^\perp \frac{\pbar_\perp}{B_0} + \pbar_\perp(\rhobbar_2 \cdot \ab_0) \right. 
 \\[1mm]
 &\qquad \left. + \pbar_\para \frac{\pbar_\perp^2}{B_0^2}\, \ab_0 \cdot \Nabla \bb_0 \cdot \cb_0  - 
\frac{1}{3} \frac{\pbar_\perp^3}{B_0^3}\cb_0 \cdot [\ab_0 \times (\ab_0 \cdot \Nabla) 
\Bb_0] \right]\ldot\thbar + \tn{rest}\,.  
\end{aligned}
\ee
From $\bb_0 \times \cb_0 = \ab_0$ it follows that the first and the third term in the 
above bracket cancel each other. In \eqref{L1} the generator $\Gbar_1^\perp$ is chosen to remove 
the Hamiltonian and the generator $\Gbar_1^\para$ is chosen to remove the parallel component 
of the vector multiplying $\ldot \Xbbar$. This leads to
\begin{align}
 \Gbar_1^\perp &= -\frac{\pbar_\para}{\pbar_\perp} \Gbar_1^\para = \frac{\pbar_\para}{\pbar_\perp} 
 \left( \Qbcal_1 \cdot \bb_0 - \frac{\pbar_\perp^2}{2 B_0}  \bb_0 \cdot \Nabla \ab_0 \cdot 
\cb_0 \right)   \label{G1perp}
 \\[1mm]
 &\ = \frac{\pbar_\para}{B_0} \left[ \frac{1}{2}\pbar_\perp \ab_0 \cdot \Nabla 
\bb_0 \cdot \cb_0 + \pbar_\para (\Nabla \times \bb_0) \cdot \cb_0 + \pbar_\perp (\Nabla 
\times \cb_0) \cdot \cb_0 - \frac{\pbar_\perp}{2}  \bb_0 \cdot \Nabla \ab_0 \cdot 
\cb_0 \right]  \nonumber
 \\[1mm]
 &\ = \frac{\pbar_\para}{B_0} \left[ -\frac{1}{2}\pbar_\perp \ab_0 \cdot \Nabla 
\bb_0 \cdot \cb_0 + \pbar_\para (\Nabla \times \bb_0) \cdot \cb_0 + \frac{\pbar_\perp}{2} 
\bb_0 \cdot \Nabla \ab_0 \cdot \cb_0 \right] \,,   \nonumber
\end{align}
Here, we used $(\Nabla \times \cb_0) \cdot \cb_0 = \bb_0 \cdot \Nabla \ab_0 \cdot \cb_0
 - \ab_0 \cdot \Nabla \bb_0 \cdot \cb_0$ to arrive at the last line. 
The first correction to the magnetic moment in \eqref{def:mubar} can now be given as the 
$\thbar$-average of $\Lbcal_2^\theta$, defined in \eqref{L2:terms},
\be \label{res:dH1}
\begin{aligned}
 -\frac{\delta H_1}{B_0} &:= \frac{1}{2\pi}\int_0^{2\pi}\dd \thbar\,\Lbcal_2^\theta = 
\frac{\pbar_\para}{B_0} \frac{\pbar_\perp^2}{2B_0} \left( - \frac{1}{2} 
(\Nabla\times\bb_0)\cdot\bb_0 + (\Nabla \ab_0 
\cdot \cb_0) \cdot \bb_0 \right) \,.
\end{aligned}
\ee
Here we used that $\Nabla\ab_0 \cdot\cb_0 = \Nabla \eb_2 \cdot \eb_1$ is independent of 
$\thbar$ as well as the average
\be
 \frac{1}{2\pi}\int_0^{2\pi}\dd \thbar\,(\ab_0 \cdot \Nabla \bb_0 \cdot \cb_0)= 
-\frac{1}{2} (\Nabla\times\bb_0)\cdot\bb_0\,.
\ee
The result \eqref{res:dH1} shows that it is straightforward to compute the inverse of 
\eqref{def:mubar} up to first order:
\be \label{inv:1st}
\frac{\pbar_\perp^2}{2} = \upmubar B_0 \left[ 1 + \eps\,\left( 
\frac{\pbar_\para}{B_0} 
\frac{1}{2} 
(\Nabla\times\bb_0)\cdot\bb_0 - \frac{\pbar_\para}{B_0} (\Nabla 
\ab_0 \cdot \cb_0) \cdot \bb_0 \right)  \right] + O(\eps^2)\,,
\ee
which corresponds to the term $\delta H_1$ stated in \eqref{delH1i}. Let us move
on to the computation of $\delta H_2$.
The term multiplying $\ldot \thbar$ in $\Lbcal_3$, denoted by $\Lbcal_3^\theta$, reads
\be
\begin{split}
 \Lbcal_3 &\: = \Lbcal_3^\theta\ldot\thbar + \tn{rest}
 \\[1mm]
 &\:=  \Bigg(- \frac{1}{2} \pder{\rhobbar_2}{\thbar} \cdot (\rhobbar_2 \times \Bb_0) + 
\Gbar_2^\perp \frac{\pbar_\perp}{B_0}  - \pder{\Gbar_1^\para}{\thbar} (\rhobbar_2 \cdot \bb_0) - 
\pder{\Gbar_1^\perp}{\thbar} (\rhobbar_2 \cdot \cb_0) + \Gbar_1^\perp (\rhobbar_2 \cdot \ab_0)  
 \\[1mm]
 &\qquad + \sum_{m=1}^2 \Gbar_{2-m}^\para\, \rhobbar_1 \cdot \Nabla \bb_0 \cdot 
\pder{\rhobbar_m}{\thbar} + 
\pbar_\para \left[\rhobbar_2 \cdot \Nabla \bb_0 + \frac{1}{2}(\rhobbar_1 \cdot \Nabla)^2 \bb_0 
\right] \cdot \frac{\pbar_\perp}{B_0}\,\cb_0 \\[1mm] 
 &\qquad + \pbar_\perp \left[ \rhobbar_1 \cdot \Nabla \cb_0 - \Gbar_1^\Theta 
\ab_0 \right]\cdot \pder{\rhobbar_2}{\thbar} + \pbar_\perp \left[ 
\frac{1}{2} \bigg(\rhobbar_1 \cdot \Nabla + \Gbar_1^\Theta \pder{}{\thbar} \bigg)^2 \cb_0 
\right] \cdot \frac{\pbar_\perp}{B_0}\,\cb_0 \\[1mm]
 &\qquad - \frac{\pbar}{B_0}\,\cb_0 \cdot [\rhobbar_2 \times (\rhobbar_1 \cdot \Nabla) 
\Bb_0] - \frac{3}{24} \frac{\pbar}{B_0}\,\cb_0 \cdot [\rhobbar_1 \times (\rhobbar_1 \cdot \Nabla)^2 
\Bb_0] \Bigg)\ldot\thbar + \tn{rest}\,.   \nonumber
\end{split}
\ee
The second correction to the magnetic moment in \eqref{def:mubar} can then be defined as the 
$\thbar$-average of $\Lbcal_3^\theta$:
\be \label{def:dH2}
\begin{aligned}
 -\frac{\delta H_2}{B_0} &:= \frac{1}{2\pi}\int_0^{2\pi} \dd\thbar\,\Lbcal_3^\theta\,.
\end{aligned}
\ee
The second-order Hamiltonian correction due to the magnetic curvature would then be the sum of 
\eqref{def:dH2} and the second-order term in \eqref{inv:1st}. We see however that such a term is 
too cumbersome for practical applications, for instance the implementation in a gyrokinetic simulation code. \\

We now proceed in the same fashion for the \emph{electrons} and substitute the polynomial 
transform \eqref{poly:transf:gc}, with $p_\perp$ 
from \eqref{def:pperp} instead of $\upmu$, in the electron starting Lagrangian \eqref{Le:start}.   
We omit again the species index for more readability. After expanding in Taylor series
the static background field, we find the following Lagrangians at the respective 
orders $\eps^n$, $0\leq n\leq 3$ (and $\Lbar_{-1} = -\Ab_0 \cdot \ldot \Xbbar$): 
\begin{align}
 \Lbar_0 &= (\sqrt{\eps}\,\pbar_\para \bb_0 + \sqrt{\eps}\,\pbar_\perp\cb_0 + \rhobbar_1 
\times \Bb_0) 
\cdot \ldot \Xbbar - \frac{\pbar_\para^2}{2} - \frac{\pbar_\perp^2}{2}\,,
 \label{L0e}
 \\[1mm]
 \Lbar_1 &= (\sqrt{\eps}\,\Gbar_1^\para \bb_0 + \sqrt{\eps}\,\Gbar_1^\perp \cb_0 + \rhobbar_2 
\times 
\Bb_0 + \Qbcal_1) \cdot \ldot \Xbbar - 
 \pbar_\para \Gbar_1^\para - \pbar_\perp \Gbar_1^\perp + \Lbcal_1\,,  
\label{L1e}
 \\[1mm]
 \Lbar_2 &= (\sqrt{\eps}\,\Gbar_2^\para \bb_0 + \sqrt{\eps}\,\Gbar_2^\perp \cb_0 + \rhobbar_3 
\times 
\Bb_0 + \Qbcal_2) \cdot \ldot \Xbbar - 
 \pbar_\para \Gbar_2^\para - \pbar_\perp \Gbar_2^\perp + \Lbcal_2\,,  
\label{L2e}
 \\[1mm]
 \Lbar_3 &= (\sqrt{\eps}\,\Gbar_3^\para \bb_0 + \sqrt{\eps}\,\Gbar_3^\perp \cb_0 + \rhobbar_4 
\times 
\Bb_0 + \Qbcal_3) \cdot \ldot \Xbbar - 
 \pbar_\para \Gbar_3^\para - \pbar_\perp \Gbar_3^\perp + \Lbcal_3\,.
\label{L3e}
\end{align}
We will then need the explicit expressions of 
\begin{equation*}
 \Qbcal_1 := -\frac{1}{2}(\rhobbar_1 \cdot \Nabla \Bb_0) \times \rhobbar_1 - 
\sqrt{\eps}\,\pbar_\para \rhobbar_1 
\times (\Nabla \times \bb_0)
- \sqrt{\eps}\,\pbar_\perp \rhobbar_1 \times (\Nabla \times \cb_0) - 
\sqrt{\eps}\,\pbar_\perp \Gbar_1^\Theta \ab_0\,, 
\end{equation*}
and of
\begin{align*}
 \Lbcal_1 := &\: \frac{1}{2} \ldot \rhobbar_1 \cdot (\rhobbar_1 \times \Bb_0) - \sqrt{\eps}\,\ldot 
\pbar_\para 
(\rhobbar_1 \cdot \bb_0) - \sqrt{\eps}\,\ldot \pbar_\perp (\rhobbar_1 \cdot \cb_0) + 
\sqrt{\eps}\,\ldot 
\thbar\,\pbar_\perp(\rhobbar_1 \cdot \ab_0)\,,
 \\[2mm]
\begin{split}
 \Lbcal_2 := &\: \ldot \rhobbar_1 \cdot (\rhobbar_2 \times \Bb_0) - \sqrt{\eps}\,\ldot \Gbar_1^\para 
(\rhobbar_1 \cdot \bb_0) - \sqrt{\eps}\,\ldot \Gbar_1^\perp (\rhobbar_1 \cdot \cb_0) + 
 \sqrt{\eps}\,\ldot \thbar\,\Gbar_1^\perp (\rhobbar_1 \cdot \ab_0)
 \\[2mm]
 & - \sqrt{\eps}\,\ldot \pbar_\para (\rhobbar_2 \cdot \bb_0) - \sqrt{\eps}\,\ldot \pbar_\perp 
(\rhobbar_2 \cdot \cb_0) + 
 \sqrt{\eps}\,\ldot \thbar\,\pbar_\perp(\rhobbar_2 \cdot \ab_0)
 \\[2mm]
 & + \sqrt{\eps}\,\pbar_\para\, \rhobbar_1 \cdot \Nabla \bb_0 \cdot \ldot \rhobbar_1 + 
 \sqrt{\eps}\,\pbar_\perp\,\rhobbar_1 \cdot \Nabla \cb_0 \cdot \ldot \rhobbar_1 - 
\sqrt{\eps}\,\pbar_\perp \Gbar_1^\Theta\,\ldot \rhobbar_1 
\cdot \ab_0
 \\[1mm]
 & + \frac{1}{3} \ldot \rhobbar_1 \cdot [\rhobbar_1 \times (\rhobbar_1 \cdot \Nabla) \Bb_0] + 
\frac{1}{6} \rhobbar_1 \times [(\rhobbar_1 \cdot \Nabla)^2 \Bb_0] \cdot \ldot \Xbbar\,.
\end{split}
\end{align*}

Here again we choose the electron generators 
$\rhob_n,\Gbar_n^\para,\Gbar_n^\perp$, and $\Gbar_n^\Theta$ in order to 
cancel as many terms as possible from the Lagrangians. In $\Lbar_0$ from \eqref{L0e} we 
require ${-\sqrt{\eps}\,\pbar_\perp\cb_0 =  \rhobbar_1 \times \Bb_0}$, which can be obtained
by setting
\be \label{rhob_e}
\rhobbar_1 := -\sqrt{\eps}\,\frac{\pbar_\perp}{B_0}\,\ab_0\,.
\ee
The factor $\sqrt\eps$ reflects the smallness of the electron Larmor radius compared to the ion 
Larmor radius. The methodology is now the same as for ions. However, the ordering and the signs of the various terms has been regarded with care. In 
\eqref{L1e} we require $\rhobbar_1 \cdot \bb_0 = 0$ and
\begin{align}
 \rhobbar_2 \times \Bb_0 &= - \sqrt{\eps}\,\Gbar_1^\perp \cb_0 - \Qbcal_1 + 
\eps\,\frac{p_\perp^2}{2 B_0} \nabla \ab_0 \cdot \cb_0 \,, \\[1mm]
 G_1^\para &= - \frac{1}{\sqrt{\eps}}\Qbcal_1 \cdot \bb_0 - \sqrt{\eps} \frac{\pbar_\perp^2}{2 B_0} 
 \bb_0 \cdot \Nabla \ab_0 \cdot \cb_0 \,,  \label{G1para:e}
 \\[1mm]
 \Gbar_1^\perp &= -\frac{\pbar_\para}{\pbar_\perp} \Gbar_1^\para\,,  \label{G1perp:e}
\end{align}
which, in contrast to \eqref{L123i}, leads to
\be \label{L1e:final}
 \Lbar_1 \sim -\eps\,\frac{p_\perp^2}{2B_0}\,\ldot{\thbar}\,.
\ee
At second order in \eqref{L2e} we can remove all terms except the gyro-averages of the terms 
multiplying $\ldot \thbar$, which in analogy to \eqref{L2:terms} are given by
\be \label{L2:terms:e}
\begin{aligned}
 \Lbcal_2 &\: = \Lbcal_2^\theta\ldot\thbar + \tn{rest}
 \\[2mm]
 &\: = \left[ -\sqrt\eps\, \frac{\pbar_\perp}{B_0}\cb_0 \cdot (\rhobbar_2 \times \Bb_0) - 
\eps\,\Gbar_1^\perp \frac{\pbar_\perp}{B_0} + \sqrt\eps\,\pbar_\perp(\rhobbar_2 \cdot \ab_0) 
\right. 
 \\[1mm]
 &\qquad \left. + \eps^{3/2} \pbar_\para \frac{\pbar_\perp^2}{B_0^2}\, \ab_0 \cdot \Nabla \bb_0 
\cdot \cb_0  - 
\frac{\eps^{3/2}}{3} \frac{\pbar_\perp^3}{B_0^3}\cb_0 \cdot [\ab_0 \times (\ab_0 \cdot \Nabla) 
\Bb_0] \right]\ldot\thbar  + \tn{rest}\,.  
\end{aligned}
\ee
Again due to $\bb_0 \times \cb_0 = \ab_0$, the first and the third term in the 
above bracket cancel each other. Moreover, from \eqref{G1para:e} and \eqref{G1perp:e} we have
\be
\begin{split}
 \Gbar_1^\perp &= -\frac{\pbar_\para}{\pbar_\perp} \Gbar_1^\para = \frac{\pbar_\para}{\pbar_\perp} 
 \left( \frac{1}{\sqrt{\eps}}\Qbcal_1 \cdot \bb_0 + \sqrt{\eps} \frac{\pbar_\perp^2}{2 B_0}  \bb_0 
\cdot \Nabla \ab_0 \cdot 
\cb_0 \right)
 \\[1mm]
 & = \sqrt{\eps}\,\frac{\pbar_\para}{B_0} \left[ -\frac{1}{2}\pbar_\perp \ab_0 \cdot \Nabla 
\bb_0 \cdot \cb_0 - \pbar_\para (\Nabla \times \bb_0) \cdot \cb_0 - \pbar_\perp (\Nabla 
\times \cb_0) \cdot \cb_0 + \frac{\pbar_\perp}{2}  \bb_0 \cdot \Nabla \ab_0 \cdot 
\cb_0 \right]
 \\[1mm]
 & = \sqrt{\eps}\, \frac{\pbar_\para}{B_0} \left[ \frac{1}{2}\pbar_\perp \ab_0 \cdot \Nabla 
\bb_0 \cdot \cb_0 - \pbar_\para (\Nabla \times \bb_0) \cdot \cb_0 - \frac{\pbar_\perp}{2} 
\bb_0 \cdot \Nabla \ab_0 \cdot \cb_0 \right] \,. 
\end{split}
\ee
Since the gyro-average of $\Lbcal_2^\theta$ is the only term that cannot be removed via the 
generators, we obtain 
\be \label{L2e:final}
 \Lbar_2 \sim \gavg{\Lbcal_2^\theta}\ldot\thbar\,,
\ee
where
\begin{equation*}
 \gavg{\Lbcal_2^\theta} 
= \frac{1}{2\pi}\int_0^{2\pi}\dd\thbar\,\Lbcal_2^\theta 
=  \eps^\threehalves \frac{\pbar_\para}{B_0} \frac{\pbar_\perp^2}{2B_0} \left[-\frac{1}{2} 
(\Nabla\times\bb_0)\cdot\bb_0 + (\Nabla \ab_0 
\cdot \cb_0) \cdot \bb_0 \right] = -\eps^\threehalves \frac{\delta H_1}{B_0}\,.  
\end{equation*}
This is the same result as for the ions in \eqref{res:dH1} but with a factor $\eps^\threehalves$.
Carrying out the analogous computations for  $\Lbar_3$, we find that generators can be chosen such 
that $\Lbar_3 = O(\eps)$. Therefore, from \eqref{L1e:final} and \eqref{L2e:final} 
we obtain the electron Lagrangian 
\be \label{Le:gc:new}
 L_\etxt \sim \left(\sqrt\eps\, \pbar_\para \bb_0 - \frac{\Ab_0}{\eps}\right) \cdot \ldot 
\Xbbar - \eps^2 \left(\frac{\pbar_\perp^2}{2 B_0} + \eps^\threehalves\frac{\delta H_1}{B_0} \right)\, \ldot 
\thbar - 
 \frac{\pbar_\para^2}{2} - \frac{\pbar_\perp^2}{2} + O(\eps^4)\,.
\ee
In contrast to 
\eqref{dtmu:i} and \eqref{def:mubar} for the ions, for the electrons we have
\be
 \dt{} \left(\frac{\pbar_\perp^2}{2 B_0} + \eps^\threehalves\frac{\delta H_1}{B_0} \right) = 
O(\eps^{2})\qquad \Rightarrow \qquad  
\upmubar := \frac{\pbar_\perp^2}{2 B_0} + \eps^\threehalves\frac{\delta H_1}{B_0}\,,
\ee
which is however less accurate than for ions, namely only up to order $O(\eps^{2})$.
The electron guiding-center Hamiltonian is obtained from the inverse of the mapping $p_\perp 
\mapsto \upmubar$:
\be \label{inv:1st}
\begin{aligned}
 &\upmubar B_0 = \frac{\pbar_\perp^2}{2} \left[ 1 + \eps^{3/2}\,\left( \frac{\pbar_\para}{B_0} 
\frac{1}{2} 
(\Nabla\times\bb_0)\cdot\bb_0 - \frac{\pbar_\para}{B_0} (\Nabla 
\ab_0 \cdot \cb_0) \cdot \bb_0 \right)  \right]
 \\[1mm]
 \Leftrightarrow \quad & \frac{\pbar_\perp^2}{2} = \upmubar B_0 \left[ 1 - \eps^{3/2}\,\left( 
\frac{\pbar_\para}{B_0} 
\frac{1}{2} 
(\Nabla\times\bb_0)\cdot\bb_0 - \frac{\pbar_\para}{B_0} (\Nabla 
\ab_0 \cdot \cb_0) \cdot \bb_0 \right)  \right] + O(\eps^2)\,,
\end{aligned}
\ee
which proves the result \eqref{Le:gc} for the electron guiding-center Hamiltonian.

\section{Gyrocenter transformation: proofs}
\label{app_proofs}
We collect here the proofs of Propositions 1-3 for ions and of Propositions 4-6
for electrons. The species index is mostly omitted for more readability. \\

{\bf Proof 1 (ion polynomial transform)}
Before proving Proposition 1, we remark that the Lagrangians $L_n$ in
\eqref{prop:i:Li}, for ${n=1,2,3}$, read
\be \label{gy:Ln}
L_n := \gab_n^{\Xb} \cdot \ldot \Xb + \gamma_n^\para\, \ldot P_\para + \gamma_n^{\upmu}\, \ldot 
\upmuhat + \gamma_n^\Theta\, \ldot \Theta - H_n \,.
\ee
The components $\gab_n^{\Xb}$, for $n=1,2,3$, are given by
\begin{subequations}
\begin{align}
 \gab_1^{\Xb} & = G_1^\para \bb_0 - \rhob_2 \times \Bb_0 + \Nabla_\perp S_2  \,, \label{gab1}
 \\[2mm]
 \gab_2^{\Xb} & = G_2^\para \bb_0 - \rhob_3 \times \Bb_0 + \Nabla_\perp S_3 + \nabla_\para S_2\, \bb_0
+ 
\Fb_\perp
 + \delta \gab_2^{\Xb} \,,
 \\[2mm]
 \gab_3^{\Xb} & = G_3^\para \bb_0 - \rhob_4 \times \Bb_0 + \nabla_\para S_3\,\bb_0 + F_\para\bb_0 + 
\delta 
\gab_3^{\Xb} 
\,, \label{gab3}
\end{align}
\end{subequations}
where the terms $\Fb_\perp$ and $\Fpar$ are defined as
\begin{subequations}
\begin{align}
\Fb_\perp & = G_1^\para\Nabla_\perp\rhob_2\cdot\bb_0-\frac{1}{2}
(\rhob_2 \times \Bb_0)\cdot\Nabla_\perp \rhob_2 + G_1^\upmu \Nabla_\perp G_1^\Theta \,, \\[1mm]
F_\para & = G_1^\para\nabla_\para\rhob_2\cdot\bb_0-\frac{1}{2}
(\rhob_2 \times \Bb_0)\cdot\nabla_\para\rhob_2+G_1^\upmu\nabla_\para G_1^\Theta \,,
\end{align}
\end{subequations}
and the terms $\delta \gab_n^{\Xb}$, for $n=2,3$, contain terms related to the curvature
of the background magnetic field and are given by
\begin{subequations}
\label{def:gamX2}
\begin{align}
\delta\gab_2^{\Xb}= & -P_\para \rhob_2 \times (\Nabla \times \bb_0) \,, \\[1mm]
\begin{split}
\delta\gab_3^{\Xb}= & -\Ppar\,\rhob_3\times(\Nabla\times\bb_0)+\Ppar\,(\rhob_3\cdot\Nabla)\bb_0 \\
& +G_1^\|(\rhob_2\cdot\Nabla)\bb_0+\onehalf(\rhob_2\cdot\Nabla\Bb_0)\times\rhob_2 \,.
\end{split}
\end{align}
\end{subequations}
The components $\gamma_n^\para$, for $n=1,2,3$, are given by
\begin{subequations}
\begin{align}
 \gamma_1^\para & = 0\,,
 \\[2mm]
 \gamma_2^\para & = - \bb_0 \cdot \rhob_2 + \pder{S_2}{P_\para} \,, \label{gampar2}
  \\
\gamma_3^\| & =-\bb_0\cdot\rhob_3+\pder{S_3}{\Ppar}
+\left(G_1^\|\bb_0-\frac{1}{2}\,\rhob_2\times\Bb_0\right)\cdot\pder{\rhob_2}{\Ppar}
+G_1^\upmu\pder{G_1^\Theta}{\Ppar} \,. \label{gampar3}
 \end{align}
\end{subequations}
The components $\gamma_n^\upmu$, for $n=1,2,3$, are given by
\begin{subequations}
\begin{align}
 \gamma_1^\upmu & = 0\,,
 \\[2mm]
  \gamma_2^\upmu & = - G_1^\Theta + \pder{S_2}{\upmuhat}\,,  \label{gammu2}
 \\
\gamma_3^\upmu & =-G_2^\Theta+\pder{S_3}{\upmuhat}
+\left(G_1^\|\bb_0-\frac{1}{2}\,\rhob_2\times\Bb_0\right)\cdot\pder{\rhob_2}{\upmuhat}
+G_1^\upmu\pder{G_1^\Theta}{\upmuhat} \,. \label{gammu3}
\end{align}
\end{subequations}
The components $\gamma_n^\Theta$, for $n=1,2,3$, are given by
\begin{subequations}
\begin{align}
  \gamma_1^\Theta & = \upmuhat\,,
  \\[2mm]
  \gamma_2^\Theta & = G_1^\upmu + \pder{S_2}{\Theta}\,, \label{gamTh2}
 \\
\gamma_3^\Theta & =G_2^\upmu+\pder{S_3}{\Theta}
+\left(G_1^\|\bb_0-\frac{1}{2}\,\rhob_2\times\Bb_0\right)\cdot\pder{\rhob_2}{\Theta}
+G_1^\upmu\pder{G_1^\Theta}{\Theta} \,. \label{gamTh3}
\end{align}
\end{subequations}
Finally, the Hamiltonians $H_n$, for $n=1,2,3$, are given by
\begin{subequations}
\begin{align}
H_1= &\: G_1^\upmu B_0 + P_\para G_1^\para + \Psi_1 + \delta H_1\,, \label{prop:i:H1}
\\[2mm]
\begin{split}
H_2= &\: G_2^\upmu B_0 + \upmuhat\, \rhob_2 \cdot \Nabla B_0 + P_\para G_2^\para + \frac{1}{2} 
(G_1^\para)^2+\bigg[(\rhobbar_2 + \rhob_2)\cdot\Nabla \\
& + (\Gbar_1^\para + G_1^\para) \der{}{P_\para} 
+  (\Gbar_1^\upmu + G_1^\upmu)\der{}{\upmuhat}+ (\Gbar_1^\Theta 
+ G_1^\Theta) \der{}{\Theta}\bigg] \Psi_1  \label{prop:i:H2}
\\
& +\left( G_1^\para\der{}{P_\para} + G_1^\upmu\der{}{\upmuhat} \right)\delta 
H_1 + \delta H_2 + \frac{1}{2} |\Ab_1|^2 + \pder{S_2}{t} \,,
\end{split}
\\[2mm]
H_3= &\: G_3^\upmu B_0 + \delta H_3 \,, \label{prop:i:H3}
\end{align}
\end{subequations}
where the generalized potential reads
\be \label{prop:i:psi}
\begin{aligned}
\Psi_\itxt\left(t,\frac{\Xb}{\eps} + \rhobbar_{1}(\Zb), P_\para, \upmuhat,\Theta\right)
= &\: \phi_1\left(t,\frac{\Xb}{\eps} + \rhobbar_{1}(\Zb) \right) - P_\para 
A_{1\para}\left(t,\frac{\Xb}{\eps} + \rhobbar_{1}(\Zb) \right) \\
 & -\sqrt{2 \upmuhat B_0(\Xb)} \,\cb_0(\Xb,\Theta) \cdot 
\Ab_{1\perp}\left(t,\frac{\Xb}{\eps} + \rhobbar_{1}(\Zb) \right) \,,
\end{aligned}
\ee
with $A_{1\para} := \Ab_1 \cdot \bb_0$ and $\Ab_{1\perp} := \bb_0 \times \Ab_1 \times \bb_0$.
Moreover, $\delta H_1$ is the curvature term introduced in \eqref{delH1i}, we do
not write the explicit expression of $\delta H_2$, and the explicit expression of
$\delta H_3$ is not relevant for our order of accuracy. \\

The results stated in Proposition 1 are obtained by substituting the gyrocenter
coordinate transformation \eqref{poly:transf:gy} into the guiding-center single-particle
Lagrangian \eqref{Li:gc} and computing its Taylor expansion in powers of $\eps$ up
to order $\eps^3$ (starting from $1/\eps$). We first denote by $\Gamma$ the symplectic
part of the guiding-center Lagrangian \eqref{Li:gc}:
\begin{equation}
\Gamma:=\left(\pbar_\|\bb_0+\frac{\Ab_0}{\eps}\right)\cdot\ldot{\Xbbar}
+\eps\,\upmubar\,\ldot{\thbar} \,.
\end{equation}
The coefficients $\Gamma_n$, for $n=-1,0,1,2,3$, of the Taylor expansion of $\Gamma$ read
\begin{subequations}
\begin{align}
\Gamma_{-1}= &\: A_0\cdot\ldot\Xb \,, \\[5pt]
\Gamma_0= &\: \Ppar\bb_0\cdot\ldot\Xb \,, \\[5pt]
\Gamma_1= &\: G_1^\|\bb_0\cdot\ldot\Xb+(\rhob_2\cdot\Nabla)\Ab_0\cdot\ldot\Xb
+\Ab_0\cdot\ldot\rhob_2+\upmuhat\,\ldot\Theta \,, \\[5pt]
\begin{split}
\Gamma_2= &\: \left[G_2^\|+\Ppar(\rhob_2\cdot\Nabla)\right]\bb_0\cdot\ldot\Xb
+(\rhob_3\cdot\Nabla)\Ab_0\cdot\ldot\Xb+\Ab_0\cdot\ldot\rhob_3 \\
& +\Ppar\bb_0\cdot\ldot\rhob_2+G_1^\upmu\ldot\Theta+\upmuhat\,\ldot G_1^\Theta \,,
\end{split} \\[5pt]
\begin{split}
\Gamma_3= &\: \left[G_3^\|+\Ppar(\rhob_3\cdot\Nabla)+G_1^\|(\rhob_2\cdot\Nabla)\right]
\bb_0\cdot\ldot\Xb
+(\rhob_4\cdot\Nabla)\Ab_0\cdot\ldot\Xb \\
& +\Ab_0\cdot\ldot\rhob_4+\frac{1}{2}(\rhob_2\cdot\Nabla)^2
\Ab_0\cdot\ldot\Xb
+\left[G_1^\|\bb_0+(\rhob_2\cdot\Nabla)\Ab_0\right]\cdot\ldot\rhob_2 \\[1.2mm]
& +\Ppar\bb_0\cdot\ldot\rhob_3+G_2^\upmu\ldot\Theta+\upmuhat\,\ldot G_2^\Theta+G_1^\upmu\ldot G_1^\Theta \,.
\end{split}
\end{align}
\end{subequations}
The results for the symplectic part follow by using the equivalence relations (for
generic generators $\rhob$ and $G^\Theta$)
\begin{subequations}
\begin{align}
(\rhob\cdot\Nabla)\Ab_0\cdot\ldot\Xb+\Ab_0\cdot\ldot\rhob & \sim
-(\rhob\times\Bb_0)\cdot\ldot\Xb \,, \\[5pt]
\Ppar\bb_0\cdot\ldot\rhob & \sim -\Ppar(\Nabla\bb_0\cdot\rhob)\cdot\ldot\Xb
-(\bb_0\cdot\rhob)\ldot\Ppar \,, \\[5pt]
\begin{split}
(\rhob\cdot\Nabla)\Ab_0\cdot\ldot\rhob+\onehalf(\rhob\cdot\Nabla)^2\Ab_0\cdot\ldot\Xb
& \sim -\onehalf(\rhob\times\Bb_0)\cdot\ldot\rhob \\
& \quad +\onehalf\left[(\rhob\cdot\Nabla\Bb_0)\times\rhob\right]\cdot\ldot\Xb \,,
\end{split} \\[5pt]
\upmuhat\,\ldot G^\Theta & \sim-G^\Theta\,\ldot{\upmuhat} \,,
\end{align}
\end{subequations}
together with the vector identity
$(\rhob\cdot\Nabla)\bb_0-\Nabla\bb_0\cdot\rhob=-\rhob\times(\Nabla\times\bb_0)$,
and by adding the terms corresponding to the total differentials $\ldot S_2$ and
$\ldot S_3$. For the Hamiltonian part, we note that the fluctuating potential $\Psi_1$
must be first transformed to the guiding-center coordinates $\Zbbar$ and then to the
preliminary gyrocenter coordinates $\Zb$. We first recall that in physical coordinates
we have
\be \label{Psi:xp}
 \Psi_1\left( t,\frac{\xb}{\eps}, p_\para,\upmu,\theta \right) = \phi_1\left(t, 
\frac{\xb}{\eps} \right) - p_\para A_{1\para}\left(t,\frac{\xb}{\eps} \right) - 
\sqrt{2\upmu\,B_0(\xb)}\, \cb_0(\xb,\theta) \cdot \Ab_{1\perp}\left(t, 
\frac{\xb}{\eps} \right)\,.
\ee
We shall first substitute the guiding-center coordinate transformation \eqref{poly:transf:gc}.
Using that $\phi_1$ and $\Ab_1$ in \eqref{Psi:xp} are normalized functions with size
and variations of order $O(1)$ in the limit $\eps\to 0$, we can safely expand in a
Taylor series around $(\Xbbar/\eps + \rhobbar_{1}, \pbar_\para,\upmubar,\thbar)$
and obtain
\begin{equation}
\label{dpsi2}
\begin{split}
\Psi_1\left(t,\frac{\xb}{\eps},\ppar,\upmu,\theta\right)= &\:
\Psi_1\left(t,\frac{\Xbbar}{\eps}+\rhobbar_{1}(\upmubar,\thbar),\pbar_\|,\upmubar,\thbar\right) \\[5pt]
& +\eps\left(\rhobbar_2\cdot\Nabla+\Gbar_1^\|\der{}{\pbar_\|}
+\Gbar_{1}^\upmu\der{}{\upmubar}+\Gbar_{1}^\Theta\der{}{\thbar}\right)
\Psi_1+O(\eps^2) \,.
\end{split}
\end{equation}
The same reasoning applies when we substitute the gyrocenter coordinate transformation
\eqref{poly:transf:gy} into \eqref{dpsi2}, yielding
\begin{equation}
\begin{split} 
\Psi_1\left(t,\frac{\xb}{\eps},\ppar,\upmu,\theta\right)= &\:
\Psi_1\left(t,\frac{\Xb}{\eps}+\rhobbar_{1}(\upmuhat,\Theta),\Ppar,\upmuhat,\Theta\right) \\[5pt]
& +\eps\bigg[(\rhobbar_2+\rhob_2)\cdot\Nabla+(\Gbar_1^\|+G_1^\|)\der{}{\Ppar}
+(\Gbar_{1}^\upmu+G_1^\upmu)\der{}{\upmuhat} \\
& \qquad +(\Gbar_{1}^\Theta+G_1^\Theta)\der{}{\Theta}\bigg]\Psi_1+O(\eps^2) \,.
\end{split}
\end{equation}
This explains the second line of the second-order Hamiltonian $H_2$ in \eqref{prop:i:H2}
as well as the expression for $\Psi_1$ given in \eqref{prop:i:psi}. \\

{\bf Proof 2 (preliminary ion gyrocenter Lagrangian)}
The components $\gab_n^{\Xb}$, for $n=1,2,3$, in \eqref{gab1}-\eqref{gab3} vanish
if and only if we set
\begin{subequations}
\begin{alignat}{2}
G_1^\para & =0 \,, \qquad
&& \rhob_{2\perp}=\frac{\bb_0}{B_0}\times\Nabla_\perp S_2 \,, \label{gy:G1para} \\[1mm]
G_2^\para & =-\nabla_\para S_2-\delta\gamma_{2\para}^{\Xb} \,, \qquad
&& \rhob_{3\perp}=\frac{\bb_0}{B_0}\times\left(\Nabla_\perp S_3+\Fb_\perp+\delta\gab_{2\perp}^{\Xb}\right) \,,
\label{gy:G2para} \\[1mm]
G_3^\para & =-\nabla_\para S_3-F_\para-\delta\gamma_{3\para}^{\Xb} \,, \qquad
&& \rhob_{4\perp}=\frac{\bb_0}{B_0}\times\delta\gab_{3\perp}^{\Xb} \,. \label{gy:G3para}
\end{alignat}
\end{subequations}
The components $\gamma_n^\para$, for $n=2,3$, in \eqref{gampar2}-\eqref{gampar3}
vanish if and only if we set
\begin{subequations}
\begin{align}
& \bb_0 \cdot \rhob_2 =\pder{S_2}{\Ppar} \,, \label{gy:r2para} \\[0mm]
& \bb_0\cdot\rhob_3=\pder{S_3}{\Ppar}
+\left(G_1^\|\bb_0-\frac{1}{2}\,\rhob_2\times\Bb_0\right)\cdot\pder{\rhob_2}{\Ppar}
+G_1^\upmu\pder{G_1^\Theta}{\Ppar} \,.
\label{gy:r3para}
\end{align}
\end{subequations}
The components $\gamma_n^\upmu$, for $n=2,3$, in \eqref{gammu2}-\eqref{gammu3} vanish
if and only if we set
\begin{subequations}
\begin{align}
& G_1^\Theta =\pder{S_2}{\upmuhat} \,, \label{gy:G1theta} \\[0mm]
& G_2^\Theta=\pder{S_3}{\upmuhat}
+\left(G_1^\|\bb_0-\frac{1}{2}\,\rhob_2\times\Bb_0\right)\cdot\pder{\rhob_2}{\upmuhat}
+G_1^\upmu\pder{G_1^\Theta}{\upmuhat} \,.
\label{gy:G2theta}
\end{align}
\end{subequations}
The Hamiltonians $H_n$, for $n=1,2,3$, in \eqref{prop:i:H1}-\eqref{prop:i:H3} vanish
if and only if we set
\begin{subequations}
\begin{align}
G_1^\upmu= &-\frac{1}{B_0}\left(\Ppar G_1^\para+\Psi_\itxt+\delta H_1\right) \,, \label{gy:G1mu} \\[1mm]
\begin{split}
G_2^\upmu= & -\frac{1}{B_0}\left\{\upmuhat\, \rhob_2 \cdot \nabla B_0 + \Ppar G_2^\para
+\frac{1}{2}(G_1^\para)^2+\bigg[(\rhobbar_2 + \rhob_2)\cdot\nabla \right. \\
&\left. + (\Gbar_1^\para + G_1^\para) 
\der{}{P_\para} + (\Gbar_1^\upmu + G_1^\upmu)\der{}{\upmuhat}+ (\Gbar_1^\theta 
+ G_1^\Theta) \der{}{\Theta}\bigg]\Psi_1\right. \label{gy:G2mu} \\
& \left. +\left(G_1^\para\der{}{\Ppar}+G_1^\upmu\der{}{\upmuhat}\right)\delta H_1
+\delta H_2+\frac{1}{2}|\Ab_1|^2+\pder{S_2}{t}\right\} \,,
\end{split} \\[1mm]
G_3^\upmu=  & -\frac{\delta H_3}{B_0} \,.
\end{align}
\end{subequations}
The only degrees of freedom left are the arbitrary scalar functions $S_2$ and $S_3$.
Since these functions must be $2\pi$-periodic in the gyro-angle $\Theta$, we cannot
eliminate $\gamma_2^\Theta$ and $\gamma_3^\Theta$, given by \eqref{gamTh2}-\eqref{gamTh3},
entirely from the Lagrangian. The reason is that the equation $\pa S_n/\pa\Theta=g$,
for a given function $g$, has $2\pi$-periodic solutions $S_n$ if and only if $\gavg g=0$,
where $\gavg g$ denotes the gyro-average of $g$ defined in \eqref{def:gavg}.
Denoting by $\wt g:=g-\gavg{g}$ the fluctuating part of $g$ (with zero gyro-average),
the dependence on the gyro-angle $\Theta$ can be removed from \eqref{gamTh2}-\eqref{gamTh3}
by setting, for $n=2,3$,
\begin{equation}
\gamma_n^\Theta=\gavg{\gamma_n^\Theta} \,,
\end{equation}
or, equivalently, by requiring that $S_n$, for $n=2,3$, satisfy the differential equations
\begin{subequations}
\label{eq:Sn}
\begin{align}
\pder{S_2}{\Theta} & =-\wt{G_1^\upmu} \,, \\
\pder{S_3}{\Theta} & =-\wt{G_2^\upmu}-\wt{G_1^\|\bb_0\cdot\pder{\rhob_2}{\Theta}}
+\wt{\frac{1}{2}\,\rhob_2\times\Bb_0\cdot\pder{\rhob_2}{\Theta}}
-\wt{G_1^\upmu\pder{G_1^\Theta}{\Theta}} \,.
\end{align}
\end{subequations}
The solutions of \eqref{eq:Sn} read (with arbitrary lower bound $\Theta_0$ of integration)
\begin{subequations}
\label{Sn}
\begin{align}
S_2(\Theta) & =S_2(\Theta_0)-\int_{\Theta_0}^\Theta\dd\Theta'\,\wt{G_1^\upmu} \,, \\
\begin{split}
S_3(\Theta) & =S_3(\Theta_0) \\
& \quad -\int_{\Theta_0}^\Theta\dd\Theta'\,\left(
\wt{G_2^\upmu}+\wt{G_1^\|\bb_0\cdot\pder{\rhob_2}{\Theta}}
-\wt{\frac{1}{2}\,\rhob_2\times\Bb_0\cdot\pder{\rhob_2}{\Theta}}
+\wt{G_1^\upmu\pder{G_1^\Theta}{\Theta}}\right) \,.
\end{split}
\end{align}
\end{subequations} \\

{\bf Proof 3 (ion gyrocenter Lagrangian)}
Before proving Proposition 3, we remark that the $\delta G_2$ in \eqref{def:dG2star}
reads
\be \label{def:delG2}
 \delta G_2:= \left\langle \left(\rhobbar_2 \cdot\Nabla + \Gbar_1^\para  
\der{}{P_\para} +  \Gbar_1^\upmu \der{}{\upmuhat}+ \Gbar_1^\Theta \der{}{\Theta}\right) 
\Psi_1 \right\rangle\,. 
\ee
The term $\delta G_2$ is linear in the fluctuating potential $\Psi_1$ and couples
to higher-order generators of the guiding-center transformation.

In order to prove our results, we first note that
computing the Euler-Lagrange equation $\pa L_\itxt/\pa \Theta - \dt{} \pa L_\itxt/\pa \ldot \Theta 
= 0$ for the preliminary gyrocenter single-particle Lagrangian \eqref{Li:prelim},
and noting that $\Obar(\eps^3)$ is independent of $\Theta$, we obtain
\be \label{dtmu:i}
\dt{}\left(\upmuhat+\eps\,\gamma_2^\Theta+\eps^2\gamma_3^\Theta \right) = O(\eps^3)\,.
\ee
Hence, the gyrocenter magnetic moment $\mu$ defined in \eqref{def:mu} is conserved
with second-order accuracy in $\eps$. Let us now compute the terms $\gavg{\gamma_2^\Theta}$
and $\gavg{\gamma_3^\Theta}$ that define the transformation $\upmuhat \mapsto \mu$
in \eqref{def:mu}. From \eqref{gamTh2} and \eqref{gy:G1mu} we have
\be \label{gam2:gavg}
 \gavg{\gamma_2^\Theta} = \gavg{G_1^\upmu} = -\frac{\gavg{\Psi_1}}{B_0} - \frac{\delta H_1}{B_0} \,,
\ee
where we used the result $G_1^\para = 0$ from \eqref{gy:G1para} and the fact that
the geometric term $\delta H_1$ does not depend on the gyro-angle. Moreover, from
\eqref{gamTh3} we have
\be \label{gam3:gavg}
 \gavg{\gamma_3^\Theta} = \gavg{G_2^\upmu} - \frac{1}{2} \gavg{ \pder{\rhob_2}{\Theta} \cdot 
(\rhob_2 \times \Bb_0)} + \gavg{ \pder{G_1^\Theta}{\Theta}G_1^\upmu }\,.
\ee
The gyro-average of $G_2^\upmu$ can be computed from \eqref{gy:G2mu}, obtaining
\begin{align}
\gavg{G_2^\upmu} &=  -\frac{1}{B_0} \gavg{  \Ppar G_2^\para + \left( \rhob_2\cdot\Nabla  
+  G_1^\upmu \der{}{\upmuhat} +  G_1^\Theta \der{}{\Theta}\right)\Psi_1 + \frac{1}{2}
\gavg{|\Ab_1|^2} } \\[1mm]
 &\quad - \frac{\gavg{G_1^\upmu}}{B_0} \der{}{\upmuhat} \delta H_1 - \frac{\delta H_2}{B_0} 
-\frac{\delta G_2}{B_0} \,,  \label{G2:gavg}
\end{align}
where we used $\gavg{\rhob_2} = 0$. In order to compute the second term on the right-hand
side of \eqref{gam3:gavg}, the generator $\rhob_2$ is determined by the function $S_2$ via
\eqref{gy:G1para} and \eqref{gy:r2para}. Omitting the arbitrary lower bound of integration
$\Theta_0$, we have
\be
S_2=-\int^\Theta\dd\Theta'\,\widetilde{G_1^\upmu}=\int^\Theta\dd\Theta'\,\frac{\widetilde{\Psi_1}}{B_0} \,,
\ee
and, recalling the functional form of $\Psi_1$ in \eqref{Psi:xp}, we obtain
\be
 \pder{S_2}{P_\para} = -\int^\Theta\dd\Theta'\,\frac{\widetilde{A_{1\para}}}{B_0}\,\,.
\ee
Therefore, the generator $\rhob_2$ reads
\be
 \rhob_2 = \frac{\bb_0}{B_0} \times \Nabla_\perp \int^\Theta \dd \Theta'\, 
 \frac{\widetilde{\Psi_1}}{B_0} - \bb_0 \int^\Theta\dd\Theta'\,\frac{\widetilde{A_{1\para}}}{B_0}\,,
\ee
which leads to 
\be \label{prop3:res1}
 - \frac 12 \gavg{\pder{\rhob_2}{\Theta} \cdot (\rhob_2 \times \Bb_0)} = - \frac 12 \gavg{
\left(\frac{\bb_0}{B_0}\times \Nabla_\perp \frac{\widetilde{\Psi_1}}{B_0}\right) \cdot 
\Nabla_\perp \int^\Theta\dd\Theta'\,\frac{\widetilde{\Psi_1}}{B_0}}\,.
\ee
In order to compute the last term in \eqref{gam3:gavg}, we get from \eqref{gy:G1theta}
\be
 G_1^\Theta = \pder{S_2}{\upmuhat} =  \int^\Theta \dd\Theta'\,\der{}{\upmuhat}\frac{\widetilde{
\Psi_1}}{B_0}\,,
\ee
yielding
\be \label{prop3:res2}
 \gavg{ \pder{G_1^\Theta}{\Theta}G_1^\upmu } = - \frac{1}{2B_0^2} \der{}{\upmuhat}
\gavg{\widetilde{\Psi_1}^2}\,.
\ee
In order to get an explicit expression for $\gavg{\gamma_3^\Theta}$, we need to compute the 
right-hand side of \eqref{G2:gavg} term by term. Using \eqref{gy:G2para} and the fact that 
$\gavg{S_2} = 0$ and $\langle \delta\gamma_{2\para}^{\Xb}\rangle = 0$ from \eqref{def:gamX2},
we find $\langle G_2^\para \rangle = 0$. The second to fourth terms in \eqref{G2:gavg} read
\begin{subequations}
\begin{align}
\begin{split}
 -\frac{1}{B_0} \gavg{\rhob_2 \cdot \Nabla \Psi_1} &= 
\gavg{\left(\frac{\bb_0}{B_0^2}\times \Nabla_\perp \widetilde{\Psi_1} \right) \cdot \Nabla_\perp 
\int^\Theta\dd\Theta'\,\frac{\widetilde{\Psi_1}}{B_0}} \\
& \quad + \frac{1}{B_0} \gavg{\nabla_\para \widetilde{\Psi_1} \int^\Theta\dd\Theta'\,\frac{\widetilde{
A_{1\para}}}{B_0}}\,,
\end{split}  \\[2mm]
 -\frac{1}{B_0} \gavg{G_1^\upmu \der{}{\upmuhat} \Psi_\itxt} &= \frac{1}{2B_0^2} 
\der{}{\upmuhat}\gavg{\Psi_\itxt^2} + \frac{1}{B_0^2} \delta H_1 \der{}{\upmuhat} \gavg{\Psi_\itxt} 
\,, \label{prop3:res3}
 \\[3mm]
 -\frac{1}{B_0} \gavg{G_1^\Theta \der{}{\Theta} \Psi_\itxt} &= \frac{1}{2B_0^2} \der{}{\upmuhat} 
\gavg{\widetilde \Psi_\itxt^2}\,,
\end{align}
\end{subequations}
where we integrated by parts in order to obtain the last equality. Substitution of
\eqref{prop3:res1} and \eqref{prop3:res2}-\eqref{prop3:res3} into \eqref{gam3:gavg} yields
\begin{equation}
\label{gam3:res}
\begin{split}
\gavg{\gamma_3^\Theta}= &\: -\frac{1}{2B_0}\gavg{|\Ab_1|^2}+\frac{1}{2B_0^2}
\der{}{\upmuhat}\gavg{\Psi_1^2} \\[5pt]
& +\frac{1}{2}\gavg{\left(\frac{\bb_0}{B_0^3}\times\Nabla_\perp\wt{\Psi_1}\right)
\cdot\Nabla_\perp\int^\Theta\dd\Theta'\,\wt{\Psi_1}} \\[5pt]
& +\frac{1}{B_0^2}\gavg{\Nabla_\|
\wt{\Psi_1}\int^\Theta\dd\Theta'\,\wt{A_{1\|}}}-\frac{\delta\Gcal_2}{B_0} \\[5pt]
& +\frac{1}{B_0^2}\der{}{\upmuhat}\left(\gavg{\Psi_1}
\delta H_1+\frac{1}{2}\delta H_1^2\right)-\frac{\delta H_2}{B_0} \,.
\end{split}
\end{equation}
The generalized magnetic moment $\mu$ can now be computed explicitly as a function of the 
fluctuating potentials from \eqref{gam2:gavg} and \eqref{gam3:res}. It remains to identify the 
gyrocenter Hamiltonian. For this purpose, we need to invert \eqref{def:mu} and substitute
the result into $H_0$. From
\be
 \mu = \upmuhat + \eps \gavg{\gamma_2^\Theta}(\upmuhat) + \eps^2 \gavg{\gamma_3^\Theta}(\upmuhat)\,,
\ee
we obtain
\be
\begin{aligned}
 \upmuhat &= \mu - \eps \gavg{\gamma_2^\Theta}(\mu - \eps \gavg{\gamma_2^\Theta}(\mu)) - \eps^2 
\gavg{\gamma_3^\Theta}(\mu) + \Obar(\eps^3)
 \\[2mm]
 &= \mu - \eps \gavg{\gamma_2^\Theta}(\mu) + \eps^2 \gavg{\gamma_2^\Theta}(\mu) \der{}{\mu} 
\gavg{\gamma_2^\Theta}(\mu) - \eps^2 
\gavg{\gamma_3^\Theta}(\mu) + \Obar(\eps^3)  \,.
 \end{aligned}
\ee
At order $O(\eps^2)$ we compute the difference
\be
 \gavg{\gamma_3^\Theta} - \frac 12 \der{}{\mu} 
\gavg{\gamma_2^\Theta}^2 = \gavg{\gamma_3^\Theta} - \frac{1}{2B_0^2} \der{}{\mu} \left( 
\gavg{\Psi_1} + \delta H_1 \right)^2\,,
\ee
and obtain, using the explicit formulas \eqref{gam2:gavg} and \eqref{gam3:res} for
$\gavg{\gamma_2^\Theta}$ and $\gavg{\gamma_3^\Theta}$,
\be
\begin{aligned}
 \upmuhat = &\: \mu+ \frac{\eps}{B_0} ( \gavg{\Psi_1} + \delta H_1 )+\frac{\eps^2}{B_0}
\Bigg( \onehalf\gavg{|\Ab_1|^2}-\frac{1}{2B_0}\der{}{\upmuhat}\gavg{\widetilde{\Psi_1}^2} \\[1mm]
& -\frac{1}{2 B_0^2}\gavg{\left(\bb_0\times \Nabla_\perp \widetilde{\Psi_1}\right)
\cdot \Nabla_\perp\int^\Theta\dd\Theta'\,\widetilde{\Psi_1}} \\[1mm]
& -\frac{1}{B_0}\gavg{\nabla_\para \widetilde{\Psi_1}\int^\Theta\dd\Theta'\,\widetilde{A_{1\para}}}
+ \delta\Gcal_2+\delta H_2\Bigg)+\Obar(\eps^3)\,,
\end{aligned}
\ee
where we used the fact that $\gavg{g}^2-\gavg{g^2}=-\gavg{\wt g^2}$, for a given
function $g(\Theta)$. Substituting this into the Hamiltonian $H_0$ completes the proof. \\

{\bf Proof 4 (electron polynomial transform)}
Before proving Proposition 4, we remark that the Lagrangians $L_{\frac{n}{2}}$,
for ${n=2,\dots,6}$, read
\be \label{e:gy:Ln}
L_{\frac{n}{2}} = \gab_{\frac{n}{2}}^{\Xb} \cdot \ldot \Xb + \gamma_{\frac{n}{2}}^\para\, \ldot P_\para
+ \gamma_{\frac{n}{2}}^{\upmu}\, \ldot \upmuhat + \gamma_{\frac{n}{2}}^\Theta\,\ldot\Theta-H_{\frac{n}{2}} \,.
\ee
The components $\gab_{\frac{n}{2}}^{\Xb}$, for $n=2,\dots,6$, are given by
\begin{subequations}
\label{gab_el}
\begin{align}
\gab_1^{\Xb} & =\rhob_2\times\Bb_0 \,, \\[5pt]
\gab_\threehalves^{\Xb} & :=G_1^\|\bb_0+\rhob_\fivehalves\times\Bb_0+\nablaperp S_\fivehalves \,, \\[5pt]
\gab_2^{\Xb} & =\rhob_3\times\Bb_0+\nablaperp S_3+\Fb_{2\perp} \,, \\[5pt]
\gab_\fivehalves^{\Xb} & :=G_2\|\bb_0+\nablapar S_\fivehalves\,\bb_0+\Fb_{\fivehalves\perp}
+\delta\gab_\fivehalves^{\Xb} \,, \\[5pt]
\gab_3^{\Xb} & =\nablapar S_3\,\bb_0+\Fb_{3\perp}+F_{3\|}\,\bb_0+\delta\gab_3^{\Xb} \,,
\end{align}
\end{subequations}
where the terms $\Fb_{\frac{n}{2}\perp}$, for $n=4,5,6$, and $F_{3\|}$ are defined as
\begin{subequations}
\begin{align}
\Fb_{2\perp} & =\onehalf(\rhob_2\times\Bb_0)\cdot\nablaperp\rhob_2
-G_\onehalf^\upmu\nablaperp G_\onehalf^\Theta \,, \\[5pt]
\begin{split}
\Fb_{\fivehalves\perp} & =G_1^\|\nablaperp\rhob_2\cdot\bb_0
-G_\onehalf^\upmu\nablaperp G_1^\Theta-G_1^\upmu\nablaperp G_\onehalf^\Theta \\
& \:\quad -(\rhob_2\cdot\Nabla)\Ab_0\cdot\nablaperp\rhob_\fivehalves
-(\rhob_\fivehalves\cdot\Nabla)\Ab_0\cdot\nablaperp\rhob_2 \,,
\end{split} \\[5pt]
\begin{split}
\Fb_{3\perp} & =G_1^\|\nablaperp\rhob_\fivehalves\cdot\bb_0
+\onehalf(\rhob_\fivehalves\times\Bb_0)\cdot\nablaperp\rhob_\fivehalves
-G_1^\upmu\nablaperp G_1^\Theta \\
& \:\quad -(\rhob_2\cdot\Nabla)\Ab_0\cdot\nablaperp\rhob_3
-(\rhob_3\cdot\Nabla)\Ab_0\cdot\nablaperp\rhob_2 \,,
\end{split} \\[5pt]
F_{3\|} & =\onehalf(\rhob_2\times\Bb_0)\cdot\nablapar\rhob_2
-G_\onehalf^\upmu\nablapar G_\onehalf^\Theta \,,
\end{align}
\end{subequations}
and the terms $\delta\gab_{\frac{n}{2}}^{\Xb}$, for $n=5,6$, contain terms related
to the curvature of the background magnetic field and are given by
\begin{subequations}
\begin{align}
\label{delta_gamma_el}
& \delta\gab_\fivehalves^{\Xb}=-\Ppar\,\rhob_2\times(\Nabla\times\bb_0) \,, \\
& \delta\gab_3^{\Xb}=-\Ppar\,\rhob_\fivehalves\times(\Nabla\times\bb_0)
-\onehalf(\rhob_2\cdot\Nabla\Bb_0)\times\rhob_2 \,.
\end{align}
\end{subequations}
The components $\gamma_{\frac{n}{2}}^\|$, for $n=2,\dots,6$, are given by
\begin{subequations}
\label{gampar_el}
\begin{align}
\gamma_1^\| & =\gamma_\threehalves^\|=\gamma_2^\|=0 \,, \\
\gamma_\fivehalves^\| & =-\bb_0\cdot\rhob_2+\pder{S_\fivehalves}{\Ppar} \,, \\
\gamma_3^\| & =-\bb_0\cdot\rhob_\fivehalves+\pder{S_3}{\Ppar}
+\frac{1}{2}(\rhob_2\times\Bb_0)\cdot\pder{\rhob_2}{\Ppar}
-G_\onehalf^\upmu\pder{G_\onehalf^\Theta}{\Ppar} \,.
\end{align}
\end{subequations}
The components $\gamma_{\frac{n}{2}}^{\upmu}$, for $n=2,\dots,6$, are given by
\begin{subequations}
\label{gammu_el}
\begin{align}
\gamma_1^\upmu & =\gamma_\threehalves^\upmu=\gamma_2^\upmu=0 \,, \\
\gamma_\fivehalves^\upmu & =G_\onehalf^\Theta+\pder{S_\fivehalves}{\upmuhat} \,, \\
\gamma_3^\upmu & =G_1^\Theta+\pder{S_3}{\upmuhat}
+\frac{1}{2}(\rhob_2\times\Bb_0)\cdot\pder{\rhob_2}{\upmuhat}
-G_\onehalf^\upmu\pder{G_\onehalf^\Theta}{\upmuhat} \,.
\end{align}
\end{subequations}
The components $\gamma_{\frac{n}{2}}^\Theta$, for $n=2,\dots,6$, are given by
\begin{subequations}
\label{gamTh_el}
\begin{align}
\gamma_1^\Theta & =\gamma_\threehalves^\Theta=0 \,, \\
\gamma_2^\Theta & =-\upmuhat \,, \\
\gamma_\fivehalves^\Theta & =-G_\onehalf^\upmu+\pder{S_\fivehalves}{\Theta} \,, \\
\gamma_3^\Theta & =-G_1^\upmu+\pder{S_3}{\Theta}
+\frac{1}{2}(\rhob_2\times\Bb_0)\cdot\pder{\rhob_2}{\Theta}
-G_\onehalf^\upmu\pder{G_\onehalf^\Theta}{\Theta} \,.
\end{align}
\end{subequations}
Finally, the Hamiltonian $H_1$ is given by
\begin{equation}
\label{prop:i:H1_el}
\begin{split}
H_1= &\: G_1^\upmu B_0+\Ppar G_1^\|-\Phi_1+\onehalf|\Ab_1|^2 \\
& +\sqrt{\frac{B_0}{2\upmuhat}}G_\onehalf^\upmu\cb_0\cdot\Ab_{1\perp}
-\sqrt{2\upmuhat B_0}G_\onehalf^\Theta\,\ab_0\cdot\Ab_{1\perp} \\
& -\sqrt{\frac{2\upmuhat}{B_0}}\ab_0\cdot\Nabla\Ab_1\cdot\left(\Ppar\bb_0+\sqrt{2\upmuhat B_0}\cb_0\right)
+O(\sqrt{\eps}) \,,
\end{split}
\end{equation}
where $\Ab_{1\perp}:=\bb_0\times\Ab_1\times\bb_0$, as before, and the fluctuating
potentials $\Phi_1$ and $\Ab_1$ are evaluated at the position $\Xb/\eps$. We remark
that the higher-order Hamiltonians $H_2$ and $H_3$ are not relevant for our order
of accuracy.

The result is obtained by substituting the gyrocenter coordinate transformation
\eqref{poly:transf:gy} into the guiding-center single-particle Lagrangian \eqref{Le:gc}
and computing its Taylor expansion in powers of $\sqrt{\eps}$ up to order $\eps^3$
(starting from $1/\eps$). We denote again by $\Gamma$ the symplectic part of the
guiding-center Lagrangian \eqref{Le:gc},
\begin{equation}
\Gamma:=\left(\sqrt{\eps}\,\pbar_\|\bb_0-\frac{\Ab_0}{\eps}\right)\cdot\ldot{\Xbbar}
-\eps^2\,\upmubar\,\ldot{\thbar} \,.
\end{equation}
The coefficients $\Gamma_{\frac{n}{2}}$, for $n=-2,\dots,6$, of the Taylor expansion of $\Gamma$ read
\begin{subequations}
\begin{align}
\Gamma_{-1} & = -A_0\cdot\ldot\Xb \,, \\[5pt]
\Gamma_{-\onehalf} & =\Gamma_0 = 0 \,, \\[5pt]
\Gamma_\onehalf & = \Ppar\bb_0\cdot\ldot\Xb \,, \\[5pt]
\Gamma_1 & = -(\rhob_2\cdot\Nabla)\Ab_0\cdot\ldot\Xb-\Ab_0\cdot\ldot\rhob_2 \,, \\[5pt]
\Gamma_\threehalves & = G_1^\|\bb_0\cdot\ldot\Xb-(\rhob_\fivehalves\cdot\Nabla)
\Ab_0\cdot\ldot\Xb-\Ab_0\cdot\ldot\rhob_\fivehalves \,, \\[5pt]
\Gamma_2 & = -(\rhob_3\cdot\Nabla)\Ab_0\cdot\ldot\Xb-\Ab_0\cdot\ldot\rhob_3-\upmuhat\,\ldot\Theta \,, \\[5pt]
\Gamma_\fivehalves & = \left(G_2^\|+\Ppar(\rhob_2\cdot\Nabla)\right)\bb_0\cdot\ldot\Xb
+\Ppar\bb_0\cdot\ldot\rhob_2-G_\onehalf^\upmu\ldot\Theta-\upmuhat\,\ldot G_\onehalf^\Theta \,, \\[5pt]
\begin{split}
\Gamma_3 & = \Ppar(\rhob_\fivehalves\cdot\Nabla)\bb_0\cdot\ldot\Xb-\frac{1}{2}
(\rhob_2\cdot\Nabla)^2\Ab_0\cdot\ldot\Xb-(\rhob_2\cdot\Nabla)\Ab_0\cdot\ldot\rhob_2 \\
& \:\quad +\Ppar\bb_0\cdot\ldot\rhob_\fivehalves-G_1^\upmu\ldot\Theta-\upmuhat\,\ldot G_1^\Theta
-G_\onehalf^\upmu\ldot G_\onehalf^\Theta \,.
\end{split}
\end{align}
\end{subequations}
The results for the symplectic part follow by using the same equivalence relations
used for ions and by adding the terms corresponding to the total
differentials $\ldot S_\fivehalves$ and $\ldot S_3$. For the Hamiltonian part, we note
that, because of the guiding-center generator (obtained in Appendix \ref{app:gc})
\eqref{rhob_e},
\be
 \rhobbar_{1\etxt} = -\sqrt{\eps}\,\sqrt{\frac{2 \upmubar}{B_0}}\, \ab_0\,,
\ee
the fluctuating potentials $\phi_1$ and $\Ab_1$ can be expanded in a Taylor series
around $\Xb/\eps$, yielding
\begin{equation}
\begin{split} 
\Phi_1\left(t,\frac{\xb}{\eps}\right)= &\:
\Phi_1\left(t,\frac{\Xbbar}{\eps}-\sqrt{\eps}\,\sqrt{\frac{2\upmubar}{B_0}}\,
\ab_0(\Xbbar,\thbar)+O(\eps)\right) \\
= &\: \Phi_1\left(t,\frac{\Xb}{\eps}\right)-\sqrt{\eps}\,\sqrt{\frac{2\upmuhat}{B_0}}\,
\ab_0(\Xb,\Theta)\cdot\Nabla\Phi_1\left(t,\frac{\Xb}{\eps}\right)+O(\eps) \,,
\end{split}
\end{equation}
and the same for $\Ab_1$. This explains the particular form of the Hamiltonian
\eqref{prop:i:H1_el}. \\

{\bf Proof 5 (preliminary electron gyrocenter Lagrangian)}
The components $\gab_{\frac{n}{2}}^{\Xb}$, for $n=2,\dots,6$, in \eqref{gab_el}
vanish if and only if we set
\begin{subequations}
\label{gy:Gpara_el}
\begin{align}
& G_1^\|:=0 \,, \\[5pt]
& G_2^\|:=-\nablapar S_\fivehalves-\delta\gamma_{\fivehalves\|}^{\Xb}
-\sqrt{\eps}\left(\nablapar S_3+F_{3\|}+\delta\gamma_{3\|}^{\Xb}\right) \,,
\end{align}
\end{subequations}
as well as
\begin{subequations}
\label{gy:rperp_el}
\begin{align}
\rhob_{2\perp} & :=0 \,, \\[5pt]
\rhob_{\fivehalves\perp} & :=-\frac{\bb_0}{B_0}\times\nablaperp S_\fivehalves \,, \\[5pt]
\rhob_{3\perp} & :=-\frac{\bb_0}{B_0}\times\bigg[\nablaperp S_3+\Fb_{2\perp}
+\sqrt{\eps}\left(\Fb_{\fivehalves\perp}+\delta\gab_{\fivehalves\perp}^{\Xb}\right)
+\eps\left(\Fb_{3\perp}+\delta\gab_{3\perp}^{\Xb}\right)\bigg] \,.
\end{align}
\end{subequations}
The components $\gamma_{\frac{n}{2}}^\|$, for $n=5,6$, in \eqref{gampar_el} vanish if and only if we set
\begin{subequations}
\label{gy:rpara_el}
\begin{align}
& \bb_0\cdot\rhob_2=\pder{S_\fivehalves}{\Ppar} \,, \\
& \bb_0\cdot\rhob_\fivehalves=\pder{S_3}{\Ppar}+\frac{1}{2}(\rhob_2\times\Bb_0)
\cdot\pder{\rhob_2}{\Ppar}-G_\onehalf^\upmu\pder{G_\onehalf^\Theta}{\Ppar} \,.
\end{align}
\end{subequations}
The components $\gamma_{\frac{n}{2}}^\upmu$, for $n=5,6$, in \eqref{gammu_el} vanish if and only if we set
\begin{subequations}
\label{gy:Gtheta_el}
\begin{align}
& G_\onehalf^\Theta=-\pder{S_\fivehalves}{\upmuhat} \,, \\
& G_1^\Theta=-\pder{S_3}{\upmuhat}-\frac{1}{2}(\rhob_2\times\Bb_0)\cdot\pder{\rhob_2}{\upmuhat}
+G_\onehalf^\upmu\pder{G_\onehalf^\Theta}{\upmuhat} \,.
\end{align}
\end{subequations}
The Hamiltonians $H_{\frac{n}{2}}$, for $n=1,2$, in \eqref{prop:i:H1_el} vanish if and only if we set
\begin{subequations}
\label{gy:Gmu_el}
\begin{align}
G_\onehalf^\upmu= & -\frac{1}{B_0}\left(\Ppar A_{1\|}
+\sqrt{2\upmuhat B_0}\,\cb_0\cdot\Ab_{1\perp}\right) \,, \\[5pt]
\begin{split}
G_1^\upmu= & -\frac{1}{B_0}\Bigg\{\Ppar G_1^\|-\Phi_1+\onehalf|\Ab_1|^2 \\
& +\sqrt{\frac{B_0}{2\upmuhat}}G_\onehalf^\upmu\cb_0\cdot\Ab_{1\perp}
-\sqrt{2\upmuhat B_0}G_\onehalf^\Theta\,\ab_0\cdot\Ab_{1\perp} \\
& -\sqrt{\frac{2\upmuhat}{B_0}}\ab_0\cdot\Nabla\Ab_1\cdot\left(\Ppar\bb_0+\sqrt{2\upmuhat B_0}\cb_0\right)
+O(\sqrt{\eps})\Bigg\} \,.
\end{split}
\end{align}
\end{subequations}
The only degrees of freedom left are the arbitrary functions $S_\fivehalves$ and $S_3$.
As for ions, the dependence on the gyro-angle $\Theta$ can be removed from \eqref{gamTh_el}
by setting, for $n=5,6$,
\begin{equation}
\gamma_{\frac{n}{2}}^\Theta=\gavg{\gamma_{\frac{n}{2}}^\Theta} \,,
\end{equation}
or, equivalently, by requiring that $S_{\frac{n}{2}}$, for $n=5,6$, satisfy the
differential equations
\begin{subequations}
\label{eq:Sn_el}
\begin{align}
\pder{S_{\frac{5}{2}}}{\Theta} & =\wt{G_\onehalf^\upmu} \,, \\
\pder{S_3}{\Theta} & =\wt{G_1^\upmu}-
\wt{\frac{1}{2}(\rhob_2\times\Bb_0)\cdot\pder{\rhob_2}{\Theta}}
+\wt{G_\onehalf^\upmu\pder{G_\onehalf^\Theta}{\Theta}} \,.
\end{align}
\end{subequations}
The solutions of \eqref{eq:Sn_el} read (with arbitrary lower bound $\Theta_0$ of integration)
\begin{subequations}
\label{Sn_el}
\begin{align}
S_{\frac{5}{2}}(\Theta) & =S_{\frac{5}{2}}(\Theta_0)+\int_{\Theta_0}^\Theta\dd\Theta'\,\wt{G_\onehalf^\upmu} \,, \\
S_3(\Theta) & =S_3(\Theta_0)
+\int_{\Theta_0}^\Theta\dd\Theta'\,\left(
\wt{G_1^\upmu}-\wt{\frac{1}{2}(\rhob_2\times\Bb_0)\cdot\pder{\rhob_2}{\Theta}}
+\wt{G_\onehalf^\upmu\pder{G_\onehalf^\Theta}{\Theta}}\right) \,.
\end{align}
\end{subequations} \\

{\bf Proof 6 (electron gyrocenter Lagrangian)}
Using that $\gavg{\gamma_2^\Theta}=-\upmuhat-\sqrt\eps\,\gavg{G_\onehalf^\upmu}$,
the Euler-Lagrange equation $\pa L_\etxt/\pa \Theta - \dt{} \pa L_\etxt/\pa \ldot \Theta=0$
for the preliminary gyrocenter single-particle Lagrangian \eqref{Le:prelim} yields
\be \label{dtmu:e}
\dt{} \left( \upmuhat + \sqrt\eps \gavg{G_\onehalf^\upmu} - \eps\gavg{\gamma_3^\Theta} \right) 
= O(\eps^\threehalves) \,.
\ee
Hence, the gyrocenter magnetic moment $\mu$ defined in \eqref{e:def:mu} is conserved
with first-order accuracy in $\eps$. Let us now compute the terms $\gavg{G_\onehalf^\upmu}$
and $\gavg{\gamma_3^\Theta}$ that define the transformation $\upmuhat \mapsto \mu$
in \eqref{e:def:mu}. From \eqref{gy:Gmu_el} we have
\begin{align}
 \gavg{G_\onehalf^\upmu} &= -\frac{1}{B_0}P_\para A_{1\para} \,,  \label{e:gavgG1}
 \\[1mm]
 \gavg{G_1^\upmu} &= -\frac{1}{B_0} \bigg( - \phi_1 + \frac{1}{2}|\Ab_1|^2 - \gavg{(\cb_0 \cdot 
\Ab_{1\perp})^2} \\
& \quad -\gavg{(\ab_0 \cdot \Ab_{1\perp})^2}- 2\upmuhat \gavg{\ab_0 \cdot \Nabla \Ab_1
\cdot  \cb_0} \bigg)  + \Obar(\sqrt\eps)\,.
\end{align}
Moreover, from \eqref{gy:Gtheta_el} we have
\be
\gavg{G_\onehalf^\upmu\pder{G_\onehalf^\Theta}{\Theta}}=-\frac{1}{B_0}\gavg{(\cb_0\cdot\Ab_{1\perp})^2}\,.
\ee
By computing the additional terms
\be
 \gavg{(\ab_0 \cdot \Ab_{1\perp})^2} = \frac 12 \left[(\eb_1 \cdot 
\Ab_{1\perp})^2 + (\eb_2 \cdot \Ab_{1\perp})^2\right] = \frac 12 |\Ab_{1\perp}|^2\,,
\ee
and
\be
\begin{split}
 \gavg{\ab_0 \cdot \Nabla \Ab_1 \cdot  \cb_0} &= \frac 12 (\eb_2 \cdot \Nabla \Ab_1 \cdot \eb_1 - 
\eb_1 \cdot \Nabla \Ab_1 \cdot \eb_2)  \\[5pt]
 &= \frac12 \eb_2 \cdot[\eb_1 \times(\Nabla\times\Ab_1)]=-\frac12 (\Nabla\times\Ab_1)\cdot\bb_0\,,
\end{split}
\ee 
we finally obtain 
\be
\label{gam3th_el_final}
  \gavg{\gamma_3^\Theta} = \frac{1}{B_0} \left[ - \phi_1 + \frac{1}{2}A_{1\para}^2 + 
\upmuhat\,(\Nabla\times\Ab_1)\cdot\bb_0\right] + \Obar(\sqrt\eps)\,,
\ee
where we used the results $G_1^\para = 0$ and $\rhb_{2\perp} = 0$ from \eqref{gy:Gpara_el}
and \eqref{gy:rperp_el}, respectively. It remains to identify the gyrocenter Hamiltonian.
For this purpose, we need to invert \eqref{e:def:mu} and substitute the result into $H_0$. 
Thanks to \eqref{e:gavgG1}, the inversion is trivial and yields
\be
 \upmuhat = \mu + \sqrt\eps\,\frac{1}{B_0} P_\para A_{1\para} + \eps 
\gavg{\gamma_3^\Theta}(\mu) + \Obar(\eps^{3/2})\,.
\ee
Substituting this into the Hamiltonian $H_0$ completes the proof.

\bibliographystyle{unsrtnat}
\bibliography{refs}

\end{document}